\documentclass[intlimits,twoside,a4paper]{article}
\usepackage{wrapfig}
\usepackage{amssymb}
\usepackage[T2A]{fontenc}
\usepackage[cp1251]{inputenc}
\usepackage{amsmath}
\usepackage{graphicx}

\usepackage{cmpj2}
\usepackage{color}

\issue{2012}{15}{2}{23003}

\doinumber{10.5488/CMP.15.23003}

\def\be{\begin{equation}}
\def\ee{\end{equation}}
\def\bea{\begin{eqnarray}}
\def\eea{\end{eqnarray}}

\def\gsim{\mathrel{\lower.65ex\hbox{$\mathop{\kern0pt\sim}\limits
   ^{\lower.55ex\hbox{$>$}}$}}}
\def\lsim{\mathrel{\lower.65ex\hbox{$\mathop{\kern0pt\sim}\limits
   ^{\lower.55ex\hbox{$<$}}$}}}

\title[Exponential approximation for repulsive Yukawa fluid]
{Simplified exponential approximation for thermodynamics of
a hard-core repulsive \\ Yukawa fluid
}

\author[S. Hlushak, A. Trokhymchuk]{S. Hlushak\refaddr{VU,IP}, A. Trokhymchuk\refaddr{IP}
}
\addresses{
\addr{VU} Department of Chemical and Biomolecular Engineering, Vanderbilt University, Nashville TN, USA
\addr{IP} Institute for Condensed Matter Physics of the National Academy of Sciences of Ukraine, \\
1 Svientsitskii Str., 79011 Lviv, Ukraine
}

\authorcopyright{S. Hlushak, A. Trokhymchuk, 2012}

\date{Received May 3, 2012, in final form May 21, 2012}

\begin{document}

\maketitle

\begin{abstract}

Exponential approximation based on the first order mean spherical approximation
(FMSA) is applied to the study of the structure and thermodynamics
of hard-core repulsive Yukawa fluids.
The proposed theory utilizes an exponential enhancement of the analytical  solution of the FMSA due to
Tang and Lu [ {J. Chem. Phys.}, 1993, \textbf{99}, 9828]
for the radial distribution function. From comparison with computer simulation data
we have shown that at low density and low temperature conditions, where original
FMSA theory fails, the FMSA-based exponential theory predicts a
significant improvement.

\pacs 01.65.Q
\end{abstract}

\section{Introduction}

Authors are pleased and honored to contribute to the festschrift dedicated to the
{sixtieth} anniversary of Orest Pizio.
Around thirty years ago one of us (AT) started his scientific career under Orest's supervision. At that time we were dealing primarily with the electroneutral mixture of charged hard spheres that are widely known as a restricted primitive model of electrolytes. Such a model incorporates both Coulomb repulsion and Coulomb attraction.
Repulsive interaction in the form of Yukawa potential, that is the subject of the present study, is quite frequently referred to as the screened Coulomb repulsion, and  is widely used in modelling various real substances in statistical physics of soft matter.
In particular, one can apply the Yukawa repulsion to the modelling of electrostatic interactions in one-component plasmas~\cite{FaroukiJCP1994},
electrical double layer interactions in charge-stabilized colloidal dispersions~\cite{MeijerJCP1997,DavoudiPRE2000,HeinenJCP2011} and micellar surfactant solutions~\cite{surfactants}, etc.
A brief but rather comprehensive review on the application of a repulsive Yukawa model in statistical physics can be found in the paper by Hopkins et al~\cite{HopkinsPRE2005}.
Some of these applications concern a pure repulsive Yukawa model, but very often the Yukawa repulsion is only a part of the potential functions used in
modelling a wide class of the so-called complex fluids.
Among others we wish to mention the core-softened fluid models, both homogeneous and those confined to a single surface, slit pore or even porous media.
The field of core-softened fluid models is intensively developing nowadays and represents one of those in which Orest Pizio and his colleagues have recently taken interest~\cite{pizioCMP2011,pizioPhyA2009,pizioJCP2009}.

Most generally, the repulsive Yukawa model is defined by the following potential function
\begin{equation}
  u_{\rm }(r) =
\frac{\gamma}{r}\exp(-z r)\,,
\label{RYpotential}
\end{equation}
where $\gamma  > 0$ is the energy (coupling) parameter that characterizes the magnitude of repulsion
while parameter $\,z\,$ controls the decay of repulsive interaction. This model has been extensively studied
by means of various methods of statistical physics, such as  numerical solutions of Ornstein-Zernike integral
equation~\cite{DavoudiPRE2000,HeinenJCP2011,CochranJCP2004}, density functional theory~\cite{TorresJCP2010,YouJPCB2005}
and computer simulations~\cite{FaroukiJCP1994,MeijerJCP1997,CochranJCP2004,YouJPCB2005,Dijrstra2003,AzharJCP2000,RobbinsJCP1988,StevensJCP1993}.
In contrast to the attractive Yukawa potential, characterized by $\,\gamma  < 0$, the repulsive Yukawa model in general does not require the introduction of an extra core repulsion at short separations.
However, looking for the analytical theory of the repulsive Yukawa fluid, attention was turned
to the Waisman's solution of the mean-spherical approximation (MSA)  for Yukawa interaction~\cite{waisman}. For reasons of the MSA solution technique, the Yukawa potential is written in a form that slightly differs from that of equation~(\ref{RYpotential}), namely,
\begin{equation}
  u(r)=\left\{
\begin{array}{lll}
\infty, & \quad r<\sigma \\
\dfrac{\varepsilon\sigma}{r}\exp(-z\left[ r-\sigma \right]), & \quad r\geqslant\sigma\,,
\end{array}
\right.
\label{HCRYpotential}
\end{equation}
where $\sigma$ is the hard-core diameter, $\,\varepsilon=\gamma\exp(-z\sigma)\,$ corresponds to the potential energy at a hard-core contact distance, $\,r=\sigma\,$. Since the  MSA solution is not sensitive to the sign of energy parameter $\,\varepsilon \,$  (or $\,\gamma\,$), it can be applicable to both repulsive and attractive Yukawa interactions.

The MSA theory for Yukawa fluids being analytical in general, requires a solution of a set of rather complex algebraic equations.
Although the original solution due to Waisman~\cite{waisman}
was later on simplified by Blum and Hoye~\cite{BlumHoye},
Cummings and Smith~\cite{CummSmith},
Ginoza~\cite{Ginoza} and finally by Henderson {et al}~\cite{Henderson1995}
and Scalise {et al}~\cite{Scalise2010},
it still invokes  a quadratic algebraic equation from which a unique physical root corresponding to the coupling parameter (prototype of the Debye screening parameter) should be deduced.
Recently, looking for further simplification of the MSA theory, Tang and Lu~\cite{TangJCP1993}  have solved Ornstein-Zernike equation by means of perturbation theory and
developed the so-called first-order mean spherical approximation (FMSA) theory.
While the FMSA is a simpler theory,
it is only slightly less accurate than the full MSA solution,
and in contrast to the latter, is ``solvable'' at any thermodynamic conditions
\cite{TangJCP2003}.
Its simplicity, however, makes it also 
attractive for an employment in different related approaches,
such as density functional theory~\cite{Tang2004fif,Tang2005fip} or augmented perturbation theory~\cite{HlushakCMP2010}, {etc.}

Similar to the MSA theory, the FMSA solution by Tang and Lu~\cite{TangJCP1993} is valid for Yukawa potential (\ref{HCRYpotential}) in both $\,\varepsilon<0\,$ and $\,\varepsilon>0\,$ cases.
However, in the present study our interest will lie  in applying the FMSA theory to the case $\,\varepsilon>0\,$ only, and we will refer to this model as the hard-core repulsive Yukawa (HCRY) fluid.
It has been already shown~\cite{CochranJCP2004,TangJCP2005} that FMSA theory is reasonably accurate for dense HCRY fluids and at high temperatures, i.e., when Yukawa repulsion is weak or dominated by a hard-core repulsion. However, the main difficulty of the MSA-like theories for repulsive potentials
concerns the appearance of negative values of the radial distribution function at short separations in the case of the growing strength of repulsive interactions.
Thus, although this has not been documented in literature, it is expected that FMSA theory will start to fail if temperature in the HCRY fluid { becomes} lower. In particular, in what follows we will show that such a deficiency of the FMSA theory becomes especially drastic at low temperature and low density conditions. It is obvious that in this region of density and temperature parameters the problems occur not only with the FMSA predictions for the radial distribution function but also for the thermodynamics of the HCRY fluid. A natural way of  fixing such a flaw in any linear theory is to employ an exponential approximation for the radial distribution function~\cite{TangAIChE1997,BarHenRMP1976,andersen1976rra,HendersonG1}.
Besides ensuring positive values of the radial distribution function at short separations, the exponential approximation  also provides a correct asymptote of the radial distribution function at large distances.
This significant improvement for the radial distribution function can be extended towards thermodynamics by means of the so-called energy route when the internal energy of the system is evaluated using its standard definition through the radial distribution function. The main purpose of the present study is to employ this strategy  for the HCRY fluid based on the existing solution of the FMSA theory for Yukawa potential.

In the following section~2 we will introduce and determine what we are referring to as the simplified exponential (SEXP) theory based on the FMSA solution due to Tang and Lu~\cite{TangJCP1993}. This will be done for both the radial distribution function and for the thermodynamics. The results and their discussions are the subject of section~3. Since the low temperature and low density states have not been studied for the HCRY fluid so far, to verify the performance of the SEXP/FMSA theory we were forced to carry out the corresponding computer simulations. Thus, the necessary details of the canonical and constant pressure ensembles Monte Carlo techniques are briefly outlined as well.  Our concluding remarks are summarized in the last section~4. To keep the paper simple and easy to understand, some auxiliary equations are moved to the appendix.

\section{Exponential theory based on the FMSA solution}

The structure and thermodynamics of the HCRY fluid, defined according to equation~(\ref{HCRYpotential}), will be studied here within the exponential approximation based on the analytical solution of the first-order mean-spherical approximation (FMSA). The complete scheme and basic expressions of the FMSA solution for the Yukawa fluid are presented in separate papers by Tang and Lu~\cite{TangJCP1993,TangJCP2005}.
Here we recall only the equations necessary for the radial distribution function.

\subsection{FMSA solution for the radial distribution function}
Like in any linear theory, the radial distribution function within the FMSA theory is represented as a sum of two terms,
\begin{equation}
  g^{\rm FMSA}\left( r \right) = g_0\left( r \right) - \beta\varepsilon g_1\left( r \right),
  \label{grFMSA}
\end{equation}
where $\,\beta=(k_{\rm B}T)^{-1}\,$ is the inverse of temperature, while $\,g_0\left(r\right)\,$
is the radial distribution function of the reference system that is a fluid of pure hard spheres, and $\,g_1\left(r\right)\,$ is the distance dependent correction term. The FMSA theory due to Tang and Lu provides us with the Laplace transform $\,G_1(s)\,$ for the correction term~\cite{TangJCP1993,TangJCP2005}
\begin{equation}
G_1(s)=\int_0^\infty rg_1(r)\re^{-sr}\rd r = \frac{ \re^{-s}}{(s+z) Q_0^2\left( s \right) Q_0^2\left( z \right)}\,,
\label{gs1}
\end{equation}
where function $\,Q_0(s)\,$ refers to the Laplace transform of the Baxter factorization function resulting from the analytical solution of the Percus-Yevick (PY) approximation for pure hard spheres~\cite{BaxterHS},
\begin{equation}
  Q_0\left(s\right)=\frac{S(s)+12\eta L(s)\re^{-s}}{(1-\eta)^2s^3}\,,
\label{QBaxter}
\end{equation}
with $\,\eta=\pi\rho\sigma^3/6\,$ being the packing fraction parameter, while $\,L(s)\,$ and $\,S(s)\,$ are polynomials. The equation for the Laplace transform, $\,G_0(s)$, of the radial distribution function
of a fluid of pure hard spheres within the framework of the PY reads
\cite{WertheimHS,BaxterHS}
\begin{eqnarray}
G_0(s)= \int_0^\infty rg_0(r)\re^{-sr}\rd r = \frac{L\left( s \right)\re^{-s}}{\left( 1-\eta \right)^2 Q_0\left( s \right)s^2}\,.
\label{gs0}
\end{eqnarray}

For the purpose of the present study it is important to have rather compact expressions in order to invert the two functions, $\,G_1(s)\,$ and $\,G_0(s)$, into a real space. In the case of pure hard spheres, it reads~\cite{SmithHS,TangMP1997}
\begin{equation}
  g_0\left(r\right)=\frac{1}{(r/\sigma)}
  \sum_{n=0}^{\infty}\left( -12\eta \right)^n C\left( 1,n+1,n+1,r/\sigma-n-1 \right)\,.
  \label{g0}
\end{equation}
A similar inversion equation for the FMSA correction term is given by~\cite{HendersonG1,TangMP1997}
\begin{equation}
  g_1\left(r\right) =
  \frac{1}{(r/\sigma)} \frac{(1-\eta)^4}{ Q_0^{2}\left( z\sigma \right)}
  \sum_{n=0}^{\infty}\left(n+1\right)\left( -12\eta \right)^n D\left( 6,n,n+2,z\sigma,\frac{r}{\sigma}-n-1 \right)\,.
  \label{g1}
\end{equation}
The functions $\,C\left( n_1,n_2,n_3,r/\sigma \right)\,$ and $\,D\left( n_1,n_2,n_3,z\sigma,r/\sigma \right)\,$ as well as the still  undefined polynomials $\,L(s)\,$ and $\,S(s)\,$ are introduced in appendix.

\subsection{SEXP/FMSA approximation for the radial distribution function}
\begin{figure}[ht]
  \centerline{
    \includegraphics[width=6.2cm]{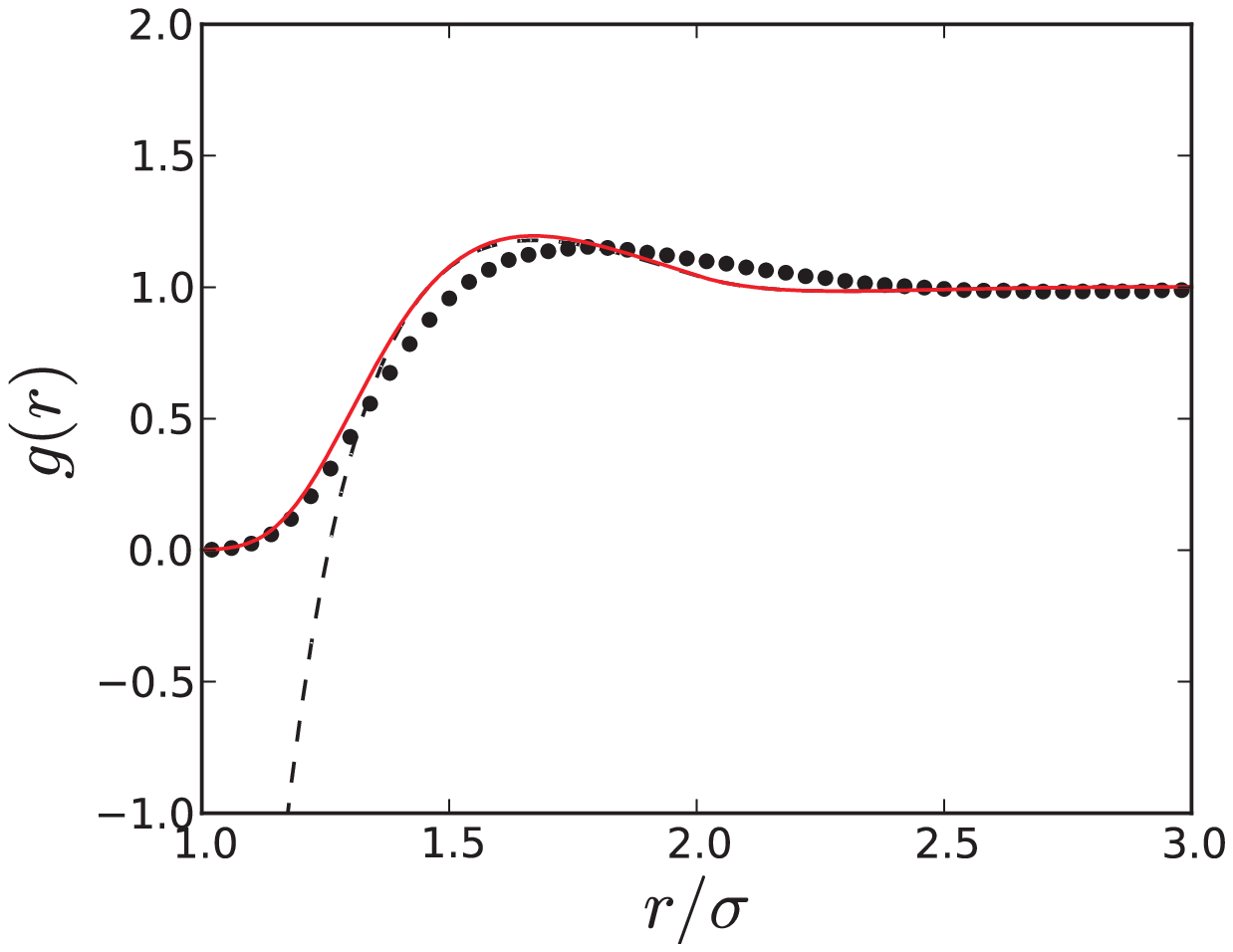}
    \includegraphics[width=6.2cm]{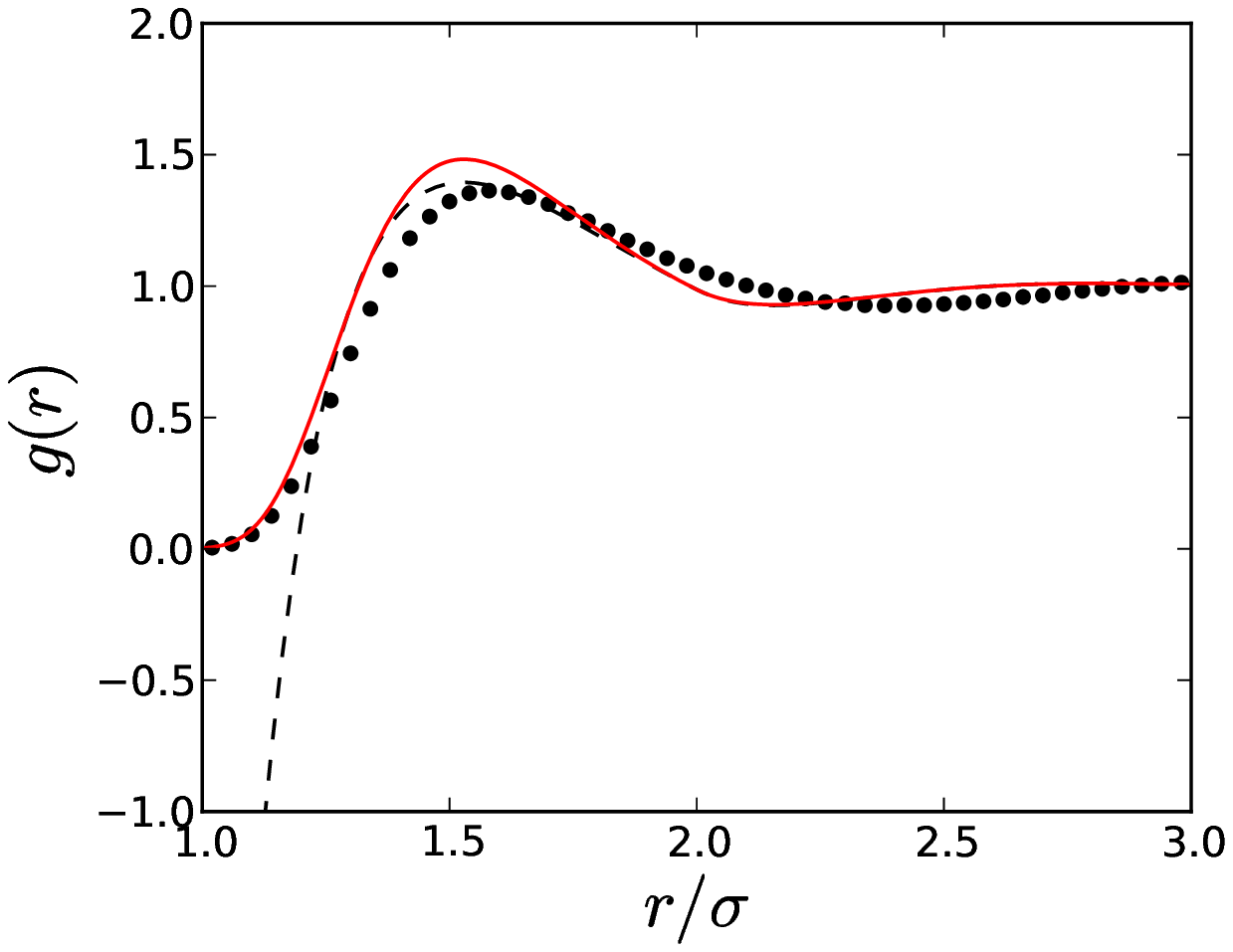}
}
\centerline{
    \includegraphics[width=6.2cm]{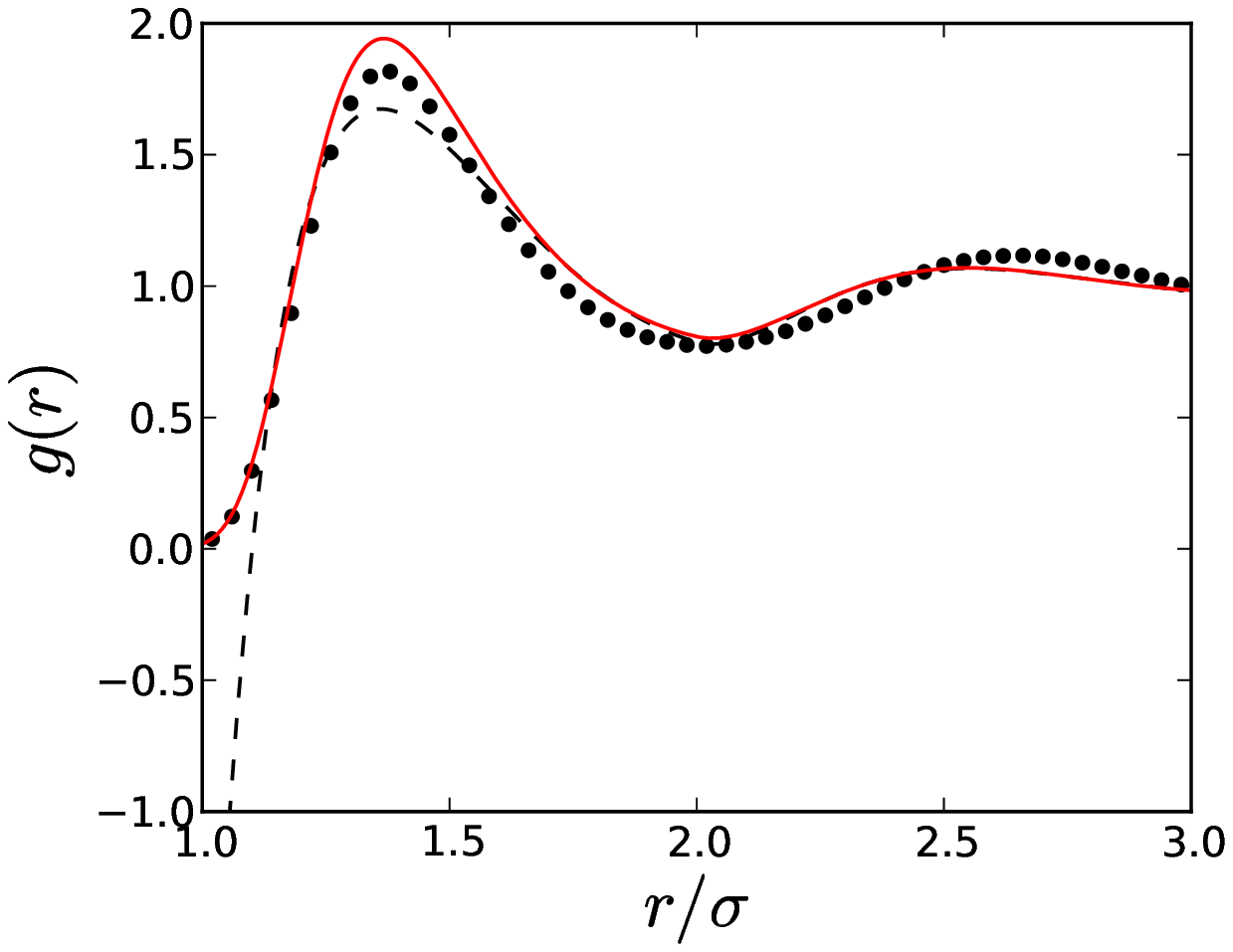}
    \includegraphics[width=6.2cm]{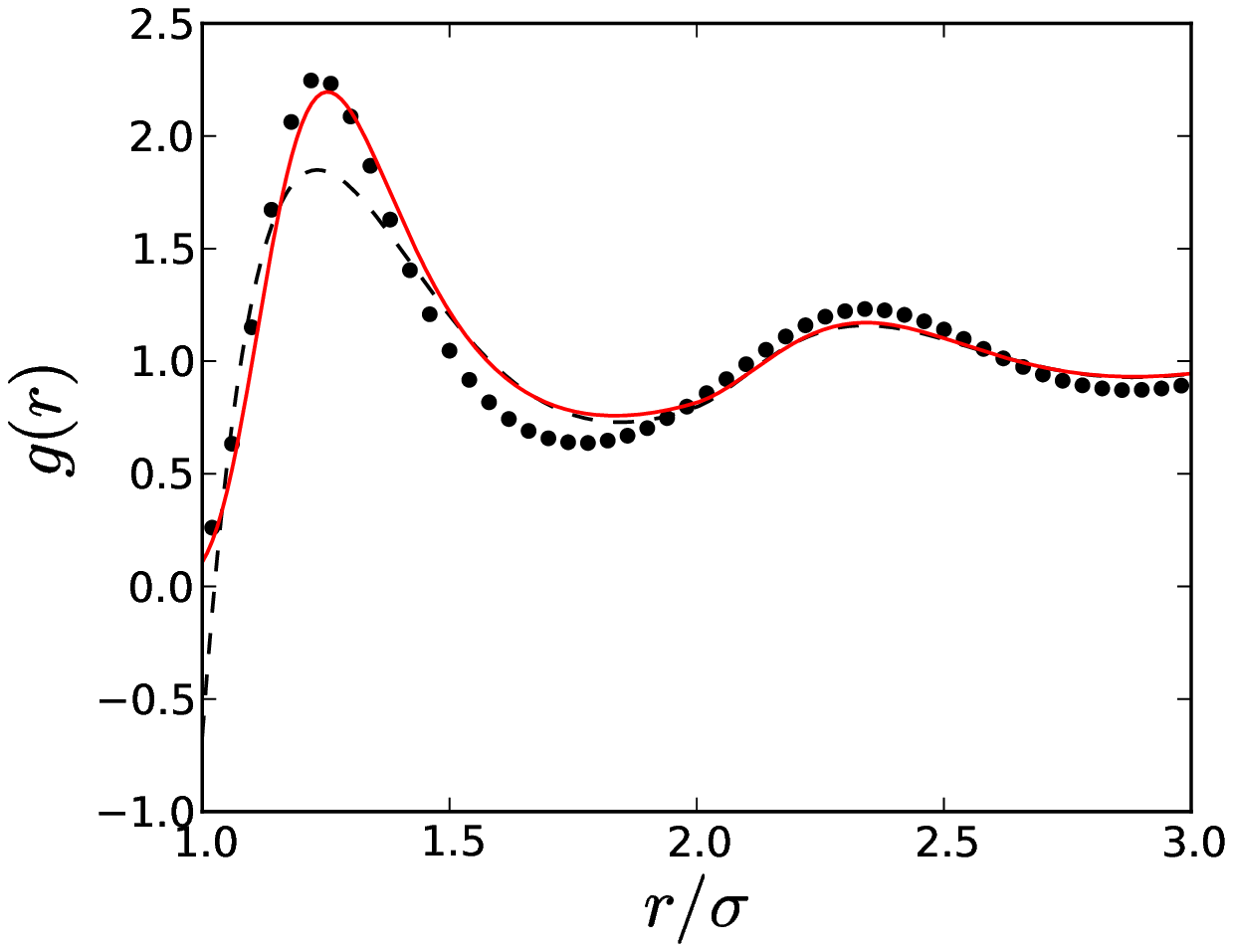}
}
  \caption{(Color online) Radial distribution functions of the hard-core repulsive Yukawa fluid
with $z\sigma=5$ at temperature $T^*=0.125$ and for several densities $\rho\sigma^3=0.1$; $0.2$; $0.4$ and $0.6$
from the left-top to the right-bottom.
Solid black lines denote the results of the FMSA theory,
while the solid red lines mark the results of the SEXP/FMSA approximation.
Symbols correspond to the MC simulation data.}
  \label{fig:z5rdf}
\end{figure}

\begin{figure}[!t]
\centerline{
 \includegraphics[width=6.2cm]{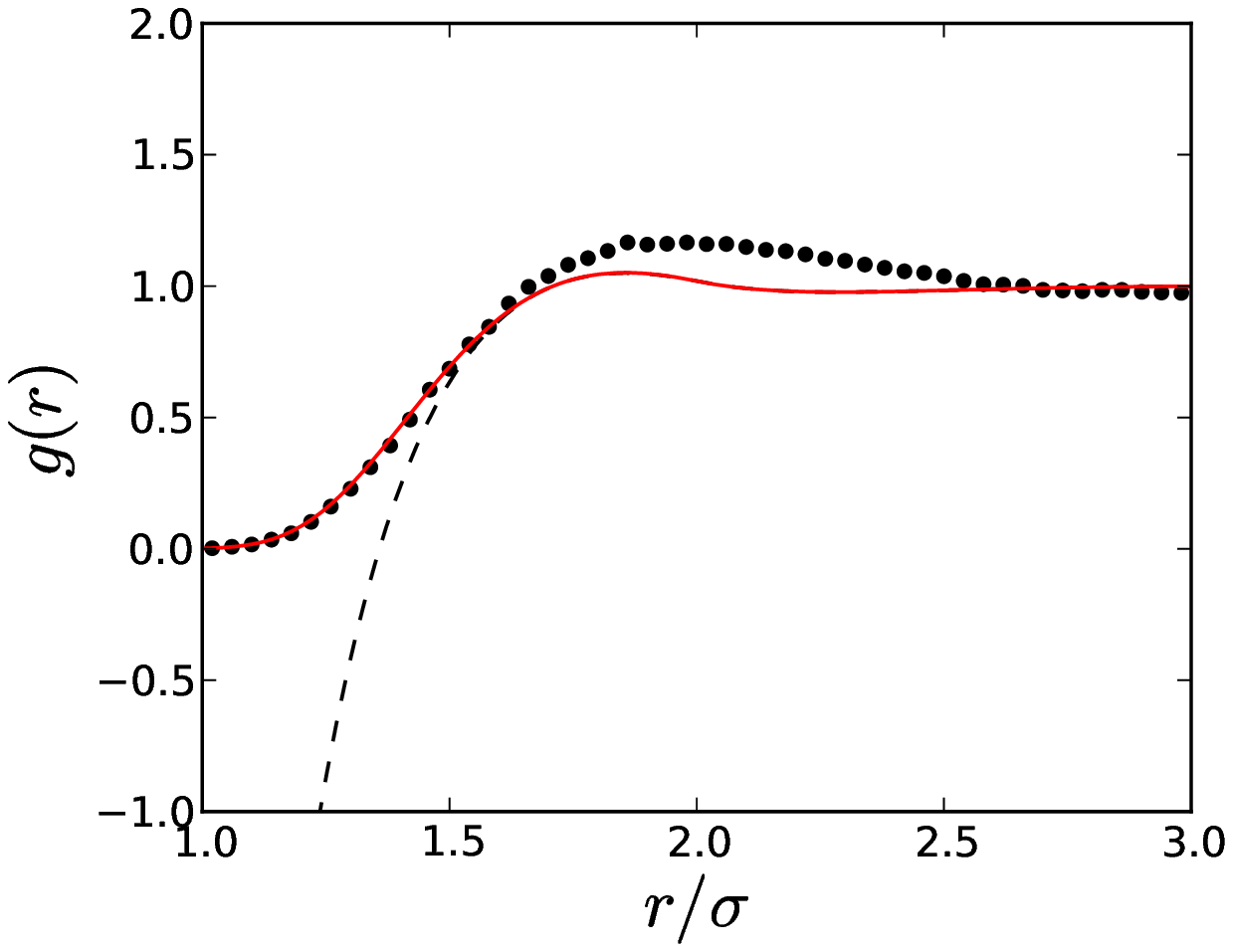}
 \includegraphics[width=6.2cm]{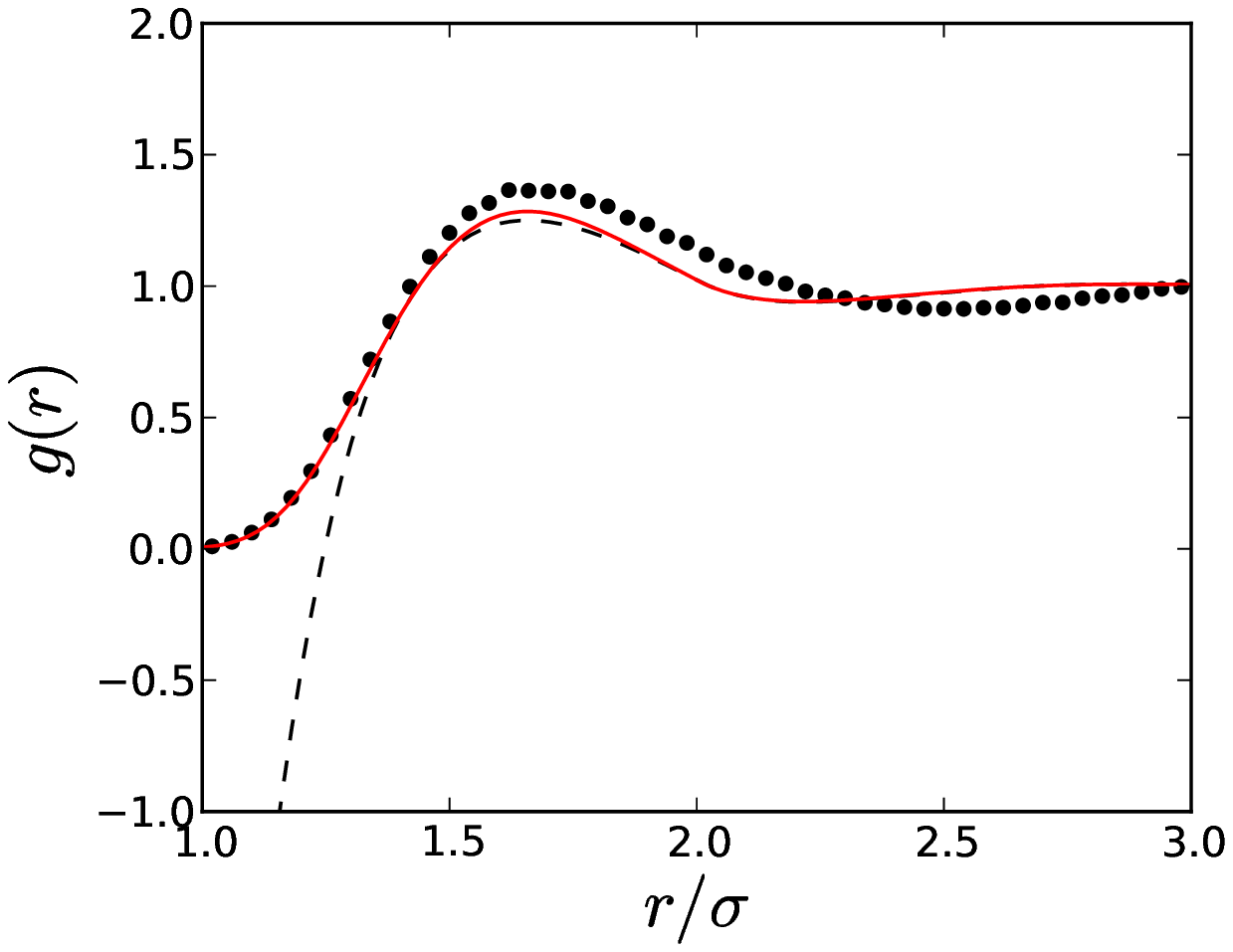}}
\centerline{
    \includegraphics[width=6.2cm]{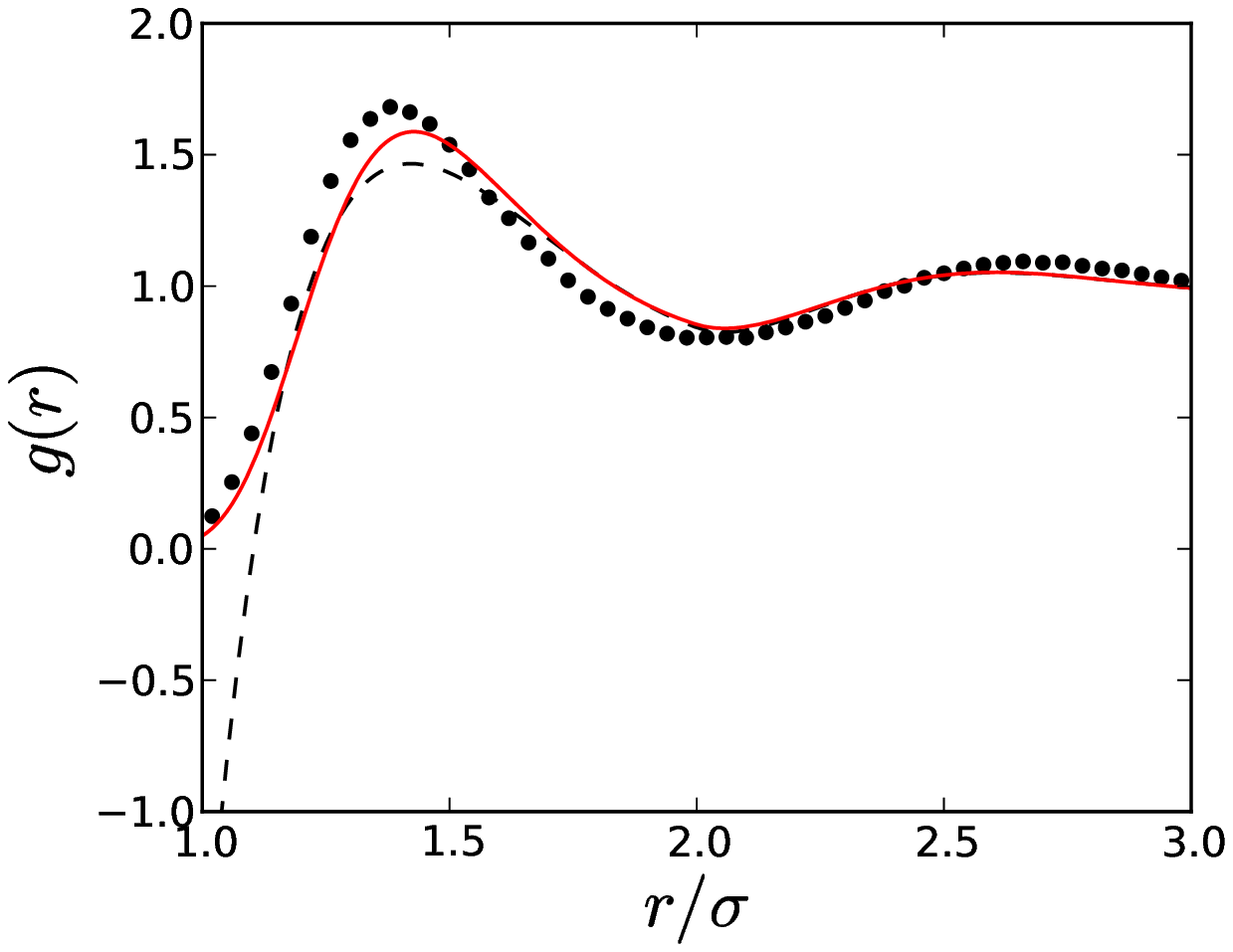}
    \includegraphics[width=6.2cm]{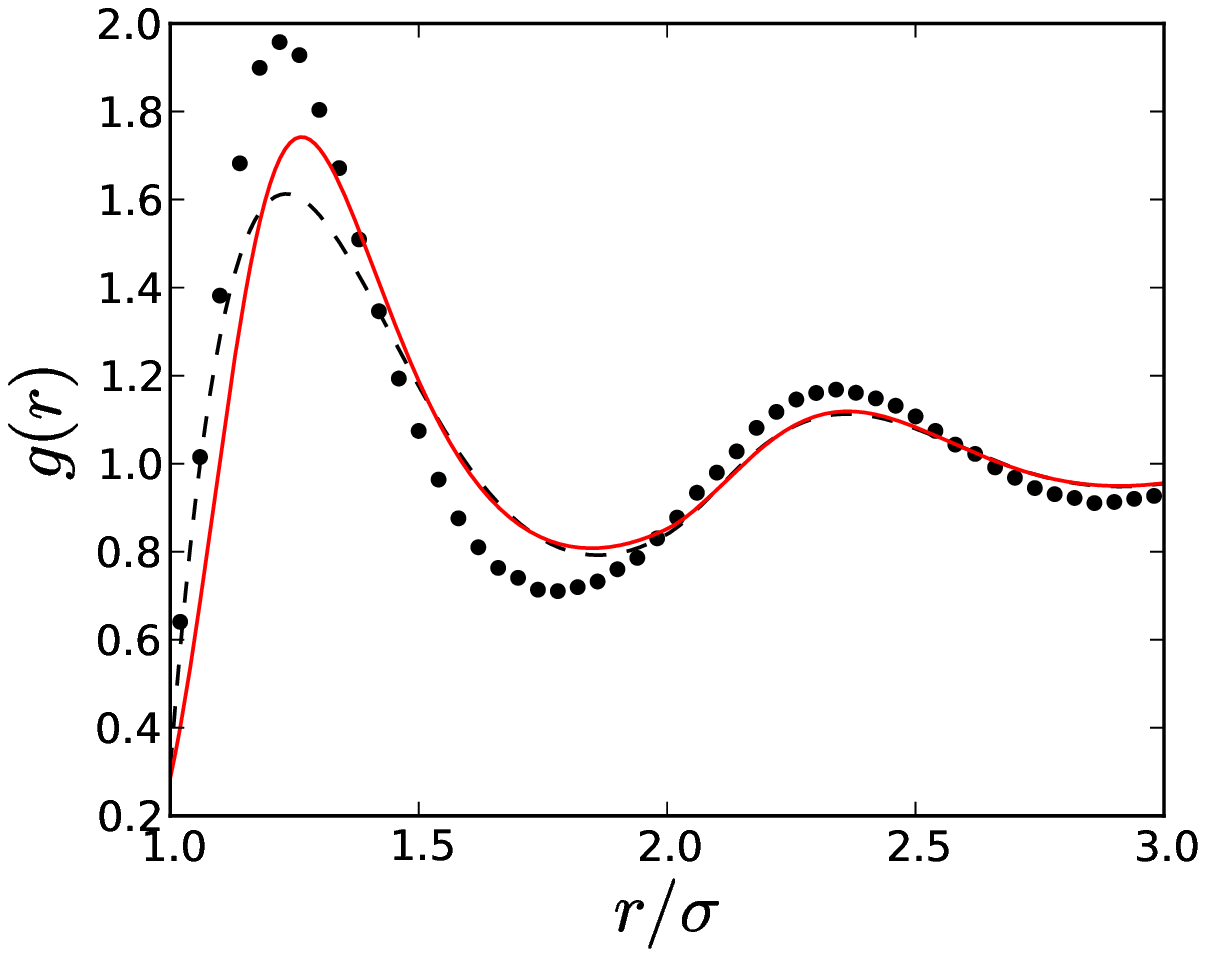}
}
  \caption{(Color online) The same as figure~\ref{fig:z5rdf} but for $z\sigma=3$.}
  \label{fig:z3rdf}
\end{figure}

\begin{figure}[!h]
  \centerline{
    \includegraphics[width=6.2cm]{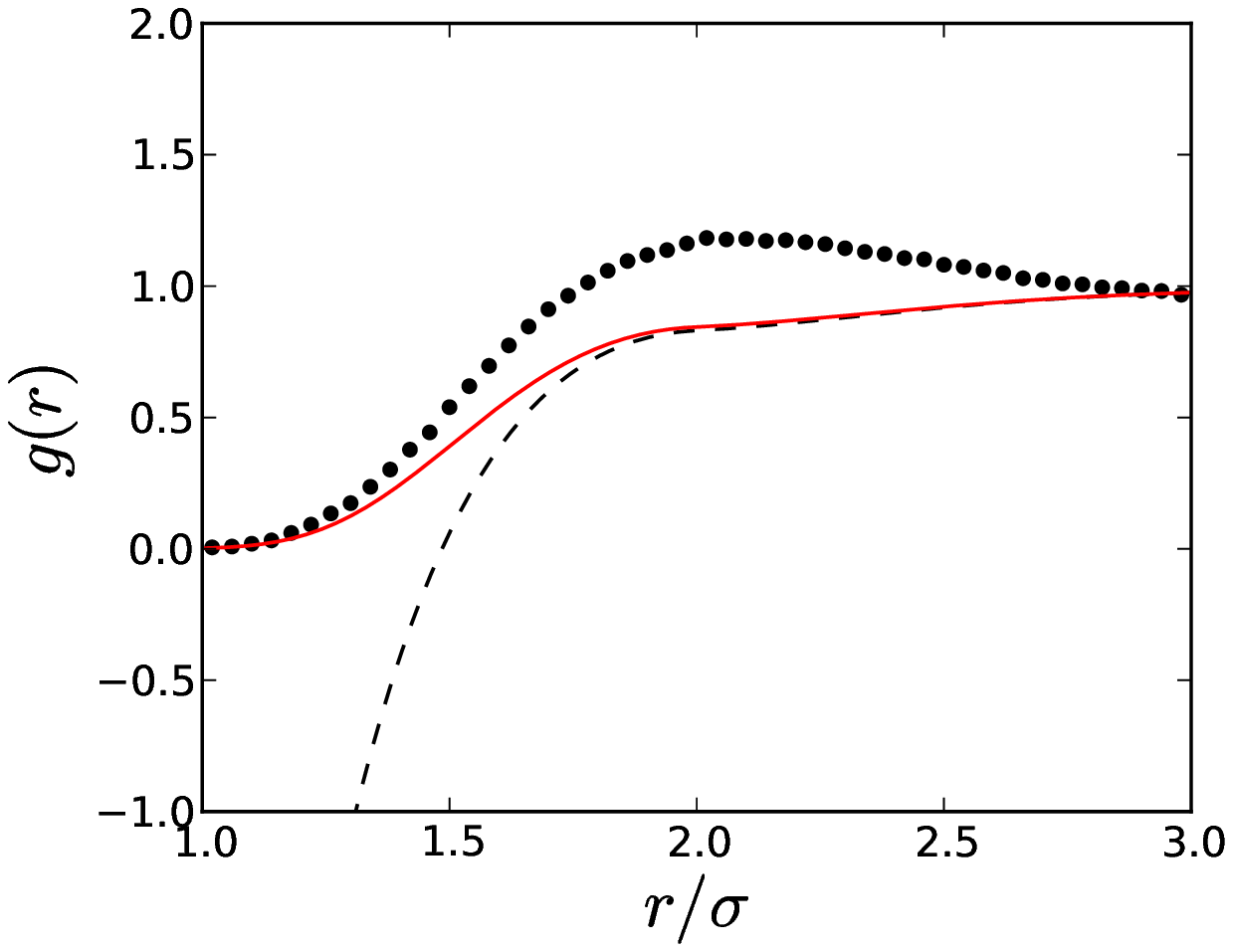}
    \includegraphics[width=6.2cm]{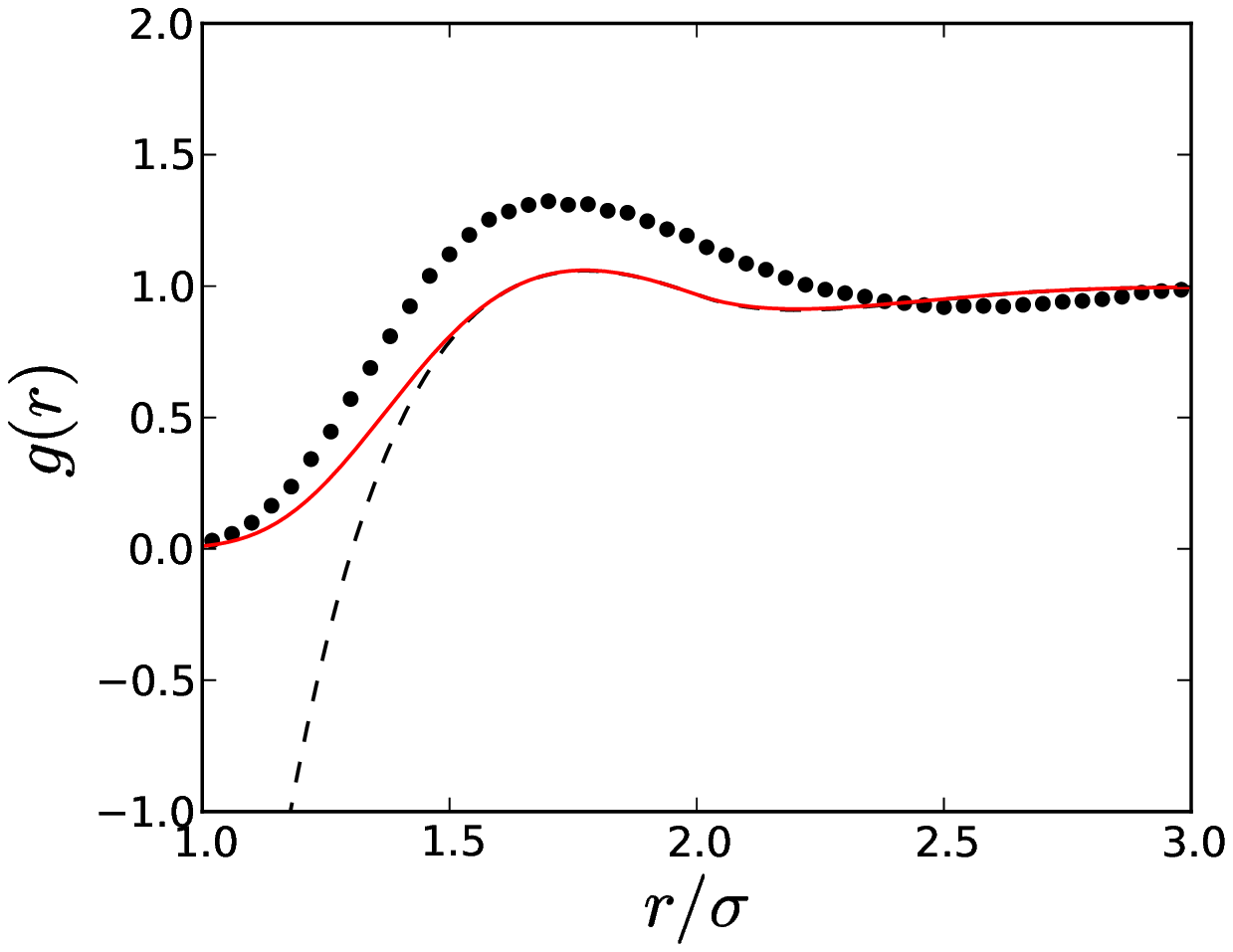}
}
\centerline{
    \includegraphics[width=6.2cm]{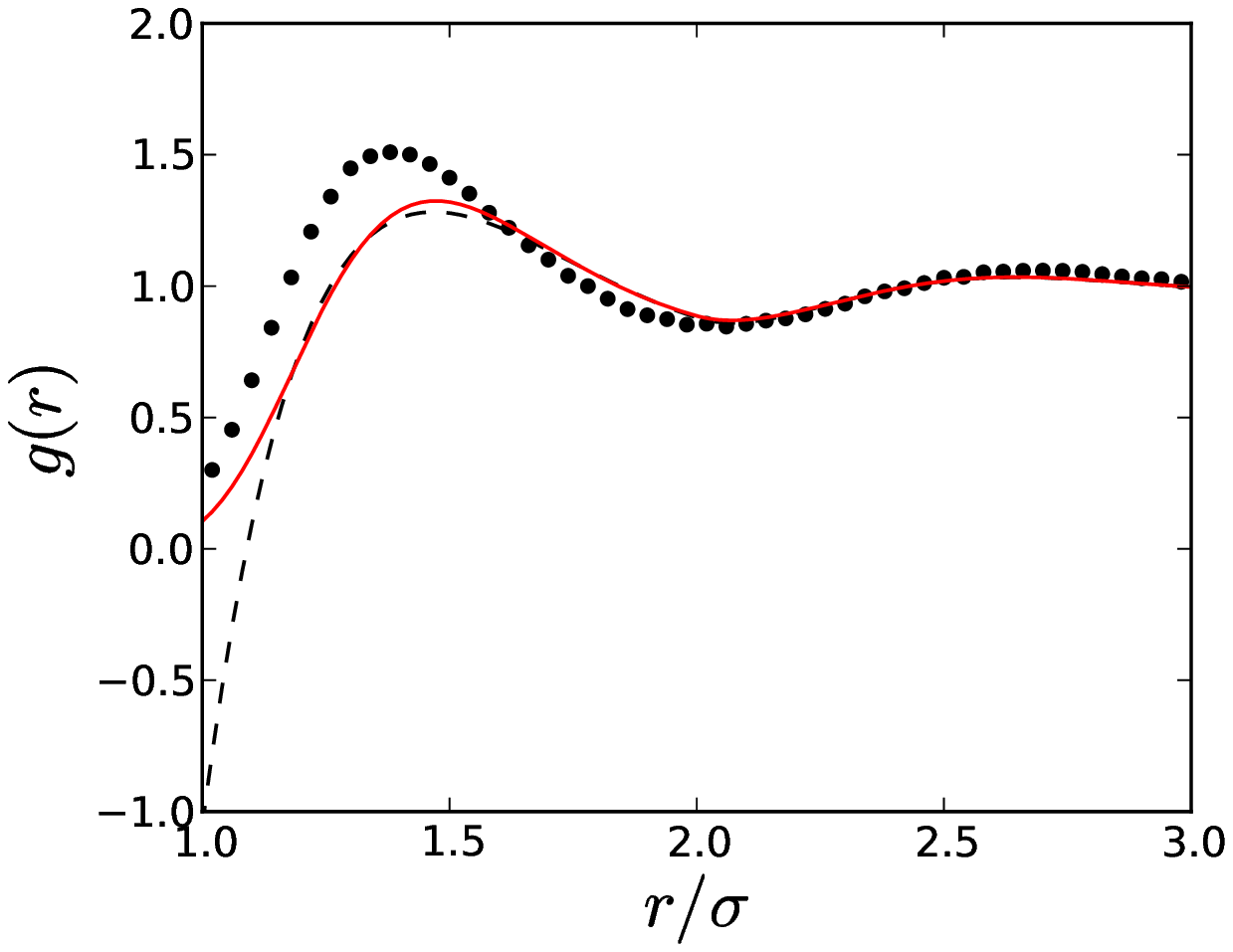}
    \includegraphics[width=6.2cm]{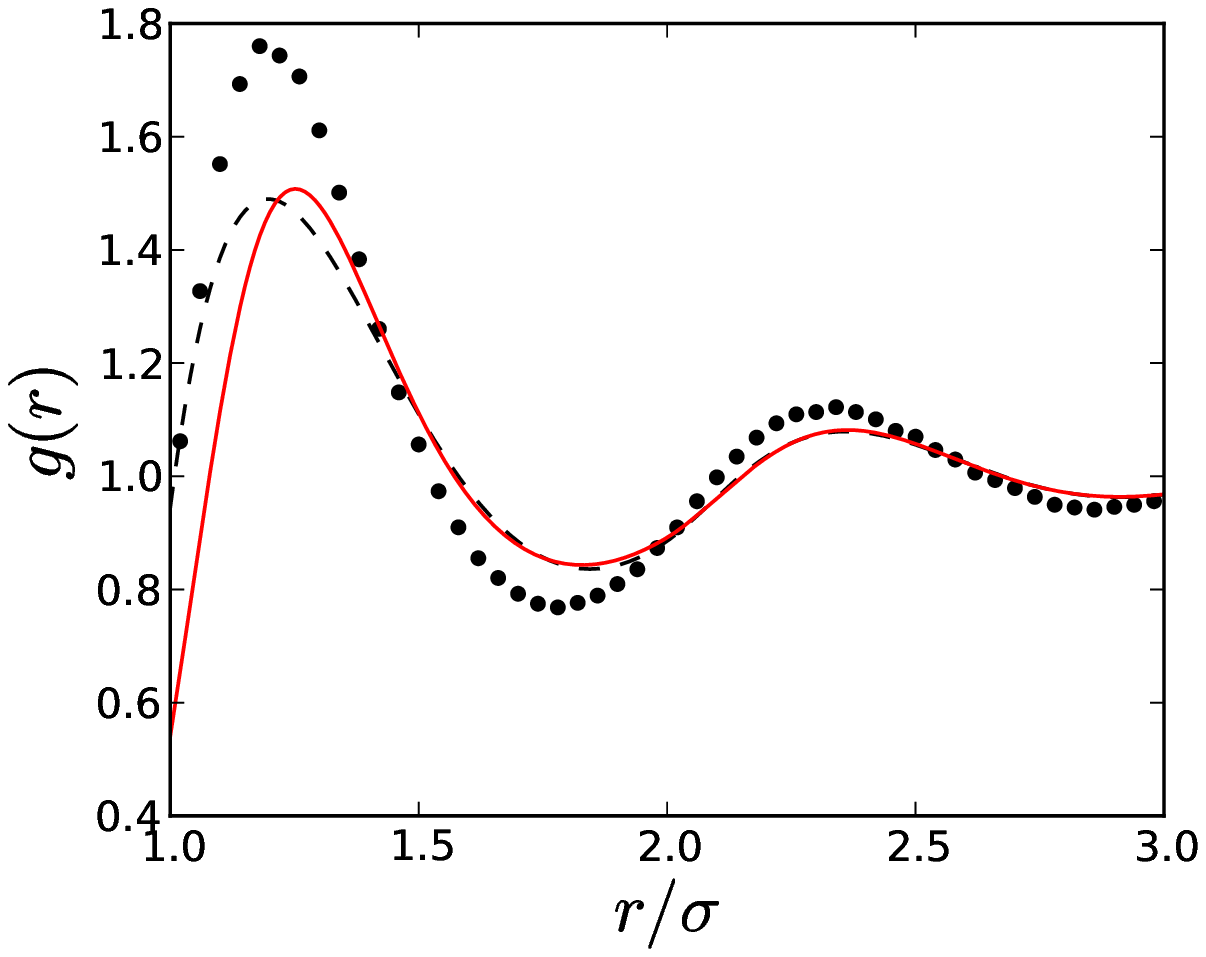}
}
  \caption{(Color online) The same as figure~\ref{fig:z5rdf} but for $z\sigma=1.8$.}
  \label{fig:z18rdf}
\end{figure}

Equations (\ref{grFMSA})--(\ref{g1}) represent the FMSA theory for the radial distribution function of the hard-core Yukawa fluid in general, i.e., for both the repulsive and  attractive types of interaction. When evaluated numerically, the results predicted by equations~(\ref{grFMSA})--(\ref{g1}) are both qualitatively and quantitatively quite similar to those obtained within the complete MSA theory
and are rather satisfactory in all practical applications  in the case of the attractive Yukawa interaction,
i.e., when $\,\varepsilon < 0\,$. However, the same FMSA theory as well as the original MSA theory for the radial distribution function are not always that efficient in the case of repulsive Yukawa interaction, i.e., for the HCRY fluid which is the subject of the present study.
In particular, by lowering the temperature for the HCRY fluid, one is actually increasing the strength of repulsive interaction.  For such conditions, the radial distribution function $\,g(r)\,$, being evaluated by means of equations~(\ref{grFMSA})--(\ref{g1}), becomes negative at short distances $\,r\,$ between particles. This is illustrated in figures~\ref{fig:z5rdf}, \ref{fig:z3rdf} and \ref{fig:z18rdf}, where the results obtained within the FMSA theory are marked by dashed lines.  Obviously, these physically incorrect results for the radial distribution function will also affect the FMSA predictions for the thermodynamics of the HCRY fluid.


A natural way of fixing such a drawback of the linear theory  is to employ an exponential (EXP) approximation for the radial distribution function as it has been suggested within the MSA theory~\cite{andersen1976rra}.  In the case of the FMSA theory,  this approximation reads
\begin{eqnarray}
  g^{\rm SEXP}\left( r \right) =  g_0\left( r \right) \exp\left[ {-\beta\varepsilon g_1\left( r \right)} \right]
  \label{gexp}\,.
\end{eqnarray}
Since FMSA theory itself is a simplified version of the more general MSA theory, the exponential approximation based on the FMSA solution is usually referred to as the simplified exponential (SEXP) approximation~\cite{TangAIChE1997}. By comparing equations (\ref{gexp}) and (\ref{grFMSA}) one can immediately conclude that $g^{\rm SEXP}\left( r \right)$ is always positive, which is due to the FMSA correction, $\,\beta\varepsilon g_1\left( r \right),\,$ residing in the exponent. These results are shown by solid lines in figures~\ref{fig:z5rdf}, \ref{fig:z3rdf} and \ref{fig:z18rdf}. Additionally, exponential approximation (\ref{gexp}) results in the radial distribution function that is asymptotically correct in the low density limit as well.
Obviously, all these improved features of the exponential approximation remain valid within the original MSA theory. The reason for  considering the SEXP/FMSA theory herein is due to the simplicity of the FMSA solution at nearly the same level of accuracy as the original MSA solution.
This circumstance has caused a growing popularity of the FMSA theory itself.  To illustrate this claim we are pointing out
that in contrast to the MSA theory, the correction function $\,g_1(r)\,$ given by equation~(\ref{g1}) does not depend on the temperature. Hence, there is very simple linear temperature dependence of the FMSA correction for the radial distribution function. This feature will be exploited herein below to write down the thermodynamics within the SEXP/FMSA theory.

\newpage
\subsection{Thermodynamics within the SEXP/FMSA approximation}
Thermodynamics within the SEXP/FMSA theory will be obtained herein through the so-called energy route~\cite{BarHenRMP1976}.
Following this scheme, the derivation of thermodynamics starts with the evaluation of the internal energy $\,E\,$ using its
definition through the radial distribution function,
\begin{equation}
  \beta\frac{ E}{N} = 2\pi\rho\beta \int_{0}^{\infty}\rd r r^2
  g^{\rm SEXP}\left( r; \beta \right)u\left( r \right)\,.
  \label{UnaN}
\end{equation}
This result is used to obtain the Helmholtz free energy in the form
\begin{equation}
  \beta\frac{A-A_{\rm id}}{N} = \beta\frac{ A_0 }{N} + \int_{0}^{\beta} \rd\beta^{\prime} \frac{E}{N} =
  \beta\frac{ A_0 }{N} + 12\eta\ \int_{\sigma}^{\infty}\rd r r^2 u\left( r \right)
  \int_{0}^{\beta} \rd\beta^{\prime} g^{\rm SEXP}\left( r; \beta^{\prime} \right)\,,
  \label{A}
\end{equation}
where $A_0$ is the Helmholtz free energy of the fluid of hard spheres.
Since functions $\,g_0\left( r \right)\,$ and $\,g_1\left( r \right)\,$
are independent of the temperature (see equation~(\ref{gexp}) for $g^{\rm SEXP}\left( r \right)$),
the inverse temperature integral in the second term of equation~(\ref{A}) can be evaluated analytically,
\begin{equation}
  \frac{1}{\beta}\int_{0}^{\beta} \rd\beta^{\prime} g^{\rm SEXP}\left( r; \beta^{\prime} \right) =
g_0\left( r \right) \exp\left[ {-\beta\varepsilon g_1\left( r \right)} \right]
\varphi_0\left[ -\beta\varepsilon g_1\left( r \right) \right]\,,
\label{bint}
\end{equation}
where we introduced notation $\,\varphi_0\left( x \right)=\left[ 1 - \exp (-x) \right]/x$.
Then, it is more convenient to rewrite the resulting Helmholtz free energy expression in the form
\begin{equation}
  \beta\frac{A-A_{\rm id}}{N} = \beta\frac{ A^{\rm FMSA}}{N} + \beta\frac{\Delta A}{N}\,,
  \label{Afinal}
\end{equation}
where we have introduced the conventional FMSA Helmholtz free energy contribution
\begin{equation}
    \beta\frac{A^{\rm FMSA}}{N} = \beta\frac{ A_0 }{N} +
  12\eta \int_{\sigma}^{\infty}\rd r r^2 u\left( r \right)
  \int_{0}^{\beta} \rd\beta^{\prime} g^{\rm FMSA}\left( r; \beta^{\prime} \right)
  = a_0 + a_1 + a_2\, ,
  \label{Afms}
\end{equation}
with coefficients $a_0$, $a_1$ and $a_2$  for the HCRY fluid given by
\cite{TangJCP2005,HendersonG1}
\begin{eqnarray}
  a_0 &=& \frac{4\eta-3\eta^2}{\left( 1-\eta \right)^2}\,,
  \label{a0} \\
  a_1 &=& 12\eta
\int_{\sigma}^{\infty}\rd r r^2 u\left( r \right)
  \int_{0}^{\beta} \rd\beta^{\prime} g_0\left( r; \beta^{\prime} \right)
  =  \frac{12\eta\varepsilon\beta L\left( z \right)}{\left( 1-\eta \right)^2Q_0\left( z \right)z^2}\,,
 \label{a1}\\
 a_2 &=& -12\eta\varepsilon
  \int_{\sigma}^{\infty}\rd r r^2 u\left( r \right)
  \int_{0}^{\beta} \rd\beta^{\prime} g_1\left( r; \beta^{\prime} \right) \beta^\prime
  = \frac{3\eta\varepsilon^2\beta^2 }{z Q_0^4\left( z \right)}\,,
  \label{a2}
\end{eqnarray}
and the correction due to the exponential approximation is
\begin{eqnarray}
\beta\frac{\Delta A}{N} = 12\eta\beta
\int_{\sigma}^{\infty}\rd r r^2 u\left( r \right)
  \left\{\frac{}{} g_0\left( r \right) \exp\left[ {-\beta\varepsilon g_1\left( r \right)} \right]
  \varphi_0\left[ -\beta\varepsilon g_1\left( r \right) \right]
- g_0\left(r\right) + \frac12\beta\varepsilon g_1\left( r \right)\right\}\,.
  \label{Aexp}
\end{eqnarray}
In contrast to equation~(\ref{Afms}), the integral in equation~(\ref{Aexp}) should be evaluated numerically by integrating along the radial distance, $\,r$, using the Gaussian quadratures.

The pressure, $\,P$, can be computed using a standard thermodynamic relation,
\begin{equation}
\beta\frac{ P}{\rho} =  \beta \mu - \beta\frac{ A}{N}\,,
\end{equation}
where $\,\mu\,$ is the chemical potential. Within the SEXP/FMSA theory, the chemical potential can be represented in the form,
\begin{eqnarray}
  \beta (\mu-\mu_{\rm id}) =  \frac{\partial}{\partial \rho}\left( \beta\frac{A-A_{\rm id}}{V} \right) =
  \frac{\partial}{\partial \rho}\left( \beta\frac{ A^{\rm FMSA}}{V} \right) + \frac{\partial}{\partial \rho}\left( \beta\frac{ \Delta A}{V} \right) =
  \beta\mu^{\rm FMSA} + \beta\Delta\mu \,.
\end{eqnarray}
This expression for chemical potential of the HCRY fluid is composed of two parts. The first part refers to
the chemical potential within the FMSA theory,
while the second one corresponds to the correction due to an exponential approximation for the radial distribution function.
These two terms can be obtained from the following equations,
\begin{equation}
 \beta\mu^{\rm FMSA} =
 \rho\left( \frac{\partial a_0}{\partial\rho}
          + \frac{\partial a_1}{\partial\rho}
          + \frac{\partial a_2}{\partial\rho}
     \right)
          + a_0 + a_1 + a_2\,,
 \label{muFMSA}
\end{equation}
\begin{eqnarray}
 \beta\Delta\mu &=&
 12\eta\beta \int_{\sigma}^{\infty}\rd r r^2 u\left( r \right)
 \left\{
 \frac{}{} g_0\left( r \right) \exp\left[ {-\beta\varepsilon g_1\left( r \right)} \right]
 \varphi_0\left[ -\beta\varepsilon g_1\left( r \right) \right]
 - g_0\left(r\right) + \frac12\beta\varepsilon g_1\left( r \right)
 \right\} \nonumber\\
&&+\quad 12\rho\eta\beta \int_{\sigma}^{\infty}\rd r r^2 u\left( r \right)
 \left\{
 \frac{}{} \frac{\partial g_0\left( r \right)}{\partial\rho}
 \exp\left[ {-\beta\varepsilon g_1\left( r \right)} \right]
 \varphi_0\left[ -\beta\varepsilon g_1\left( r \right) \right] \right. \nonumber\\
&&+\quad\left.\beta\varepsilon\frac{\partial g_1\left( r \right)}{\partial\rho}
 g_0\left( r \right) \exp\left[ {-\beta\varepsilon g_1\left( r \right)} \right]
 \varphi_1\left[ -\beta\varepsilon g_1\left( r \right) \right]
 %
- \frac{\partial g_0\left(r\right)}{\partial\rho}
 + \frac12\beta\varepsilon{\frac{\partial g_1\left( r \right)}{\partial\rho}}
 \right\} \,,
\end{eqnarray}
where $\varphi_1\left(x\right)=\left[ 1-x-\exp\left( -x \right) \right]/x^2$, and five density derivatives were introduced, i.e.,
$\,{\partial a_0}/{\partial\rho}$, $\,{\partial a_1}/{\partial\rho}$, $\,{\partial a_2}/{\partial\rho}$, $\,{\partial g_0\left(r\right)}/{\partial\rho}\,$
and $\,{\partial g_1\left(r\right)}/{\partial\rho}\,$.
The first two of them
could be straightforwardly evaluated using definitions (\ref{a0})--(\ref{a2}) for the quantities $\,a_0$, $\,a_1\,$ and $\,a_2$, respectively.
As for the next two, in accordance with equations~(\ref{g0}) and (\ref{g1}), one will be
required to evaluate the density derivatives of the functions
$\,C\left( n_1,n_2,n_3,x \right)\,$ and $\,D\left( n_1,n_2,n_3,y,x \right)$.
These derivatives are also straightforward and the details can be found in the appendix.

\section{Results and discussions}

We considered three models for the HCRY fluids with a fixed strength of repulsive energy, $\varepsilon=1$, but three different ranges of repulsive interaction, i.e., characterized by three different values of the decay parameter, namely, $\,z\sigma=5$,  $\,3\,$ and $\,1.8$. The reduced temperature was fixed at $T^*=k_{\rm B}T/\varepsilon=0.125$. While three values of the decay parameter $\,z\,$ have been already used in related studies~\cite{CochranJCP2004,TangJCP2005} of the HCRY fluids, such a low value of reduced temperature has not been explored so far in theoretical studies of this type of model fluids. At the same time, the reduced temperature $T^*=0.125$ and even lower temperatures are quite obvious in computer simulation studies of the phase diagrams of the HCRY fluid~\cite{HeinenJCP2011,AzharJCP2000}.  While considering such a low temperature, our main intention, on the one hand, is to clearly illustrate the problems experienced by linear theories in the description of purely repulsive fluids, and, on the other hand, to show an extent to which the application of exponential theories can improve the theoretical treatment of repulsive fluids.

To test the theoretical predictions,
we performed canonical ensemble and constant pressure ensemble Monte-Carlo (MC) simulations. To this end, the modified codes from the book of Frenkel and Smit~\cite{Frenkel96} were used and the radial distribution function $\,g(r)\,$, the compressibility factor $\,Z=\beta P/\rho\,$  and the internal energy $\,\beta E\,$ were calculated.
In particular, the radial distribution function was obtained from canonical ensemble MC simulations for 256 particles in a cubic simulation box with periodic boundary conditions.
All these simulations were performed for 500 equilibration steps and 1500 production steps. Each step consisted of 2500  attempts to displace a particle.
On the other hand, constant pressure ensemble MC simulations were used to sample the density dependence of the pressure and internal energy.
This set of simulations was performed for 256 particles with 2500 equilibration steps and 5000 production steps.
Similarly, each step consisted of 10000 attempts to displace a particle and 100 attempts to change the box volume.

First, for each of three considered HCRY fluids
we compare the radial distribution functions evaluated within the FMSA and SEXP/FMSA theories against MC simulations data.
These results  are shown in figures from \ref{fig:z5rdf}
to \ref{fig:z18rdf}
for the HCRY fluids characterized by $z\sigma=5$, $3$ and $1.8$, respectively. Each part of these figures consists of a set of two theoretical (lines) and one computer simulation (symbols) results for one of four densities, namely, $\,\rho\sigma^3=0.1$, 0.2, 0.4 and 0.6. As it was already mentioned, the main problem of the radial distribution functions evaluated within the linear theories, such as MSA and FMSA, concerns an appearance of the non-physical negative values of $\,g(r)\,$ at short distances $\,r\,$,
when the strength of repulsion is increasing.
Indeed, as we can see from figures~\ref{fig:z5rdf}, \ref{fig:z3rdf} and \ref{fig:z18rdf}, this is the case at the considered reduced temperature $T^*=0.125$, where radial distribution functions shown by dashed lines  (i.e., those that correspond to the FMSA theory)
lie in the region of negative values. This region of negative values of $\,g(r)\,$ is more notable and extends far beyond the contact distance $\,r/\sigma =1\,$ for lower densities $\,\rho\sigma^3=0.1$ and 0.2. We note that for these two densities, the distances $\,r\,$ for which $\,g(r)\,$ assume negative values, increase with an increase of the range of repulsion, i.e., going from $\,z\sigma=5$, to $\,z\sigma=3$ and 1.8. However, for density $\,\rho\sigma^3=0.4\,$ and larger densities, the tendency  changes to the opposite. Namely, the region of negative values of  $\,g(r)\,$ of the HCRY fluid characterized by $\,z\sigma=5\,$ (see figure~\ref{fig:z5rdf}), starts to shrink going to fluids with $\,z\sigma=3\,$ and 1.8. In the case of the highest density explored in this study, $\,\rho\sigma^3=0.6$, the radial distribution function $\,g(r)\,$, evaluated from the FMSA theory, becomes even fully positive for $\,z\sigma=3\,$ and 1.8 (see figures~\ref{fig:z3rdf} and \ref{fig:z18rdf}), while it remains always negative for the case of $\,z\sigma=5$.

\begin{table}[ht]
\caption{Contact values of the radial distribution function of the HCRY fluids with a different range of repulsive interaction
characterized by different values of the decay parameter $\,z$. Reduced temperature
is fixed at $\,T^*=0.125$.}
\vspace{2ex}
\begin{tabular*}{\textwidth}{@{\extracolsep{\fill}}rrrr|rrr|rrr}
\hline
\hline
& \multicolumn{3}{c}{$z\sigma=5$} & \multicolumn{3}{c}{$z\sigma=3$} & \multicolumn{3}{c}{$z\sigma=1.8$} \\
\cline{2-4}\cline{5-7}\cline{8-10}
$\rho\sigma^3$  &  MC  &  FMSA  &  SEXP  &  MC  &  FMSA  &  SEXP &  MC  &  FMSA  &  SEXP \\
\hline
0.1  &  0.003  &  --5.943   &  0.001 & 0.001 &  --5.549  &  0.001&  0.007 &  --5.154   &  0.002 \\
0.2  &  0.006  &  --4.891   &  0.003 & 0.008 &  --4.209  &  0.005&  0.025 &  --3.580   &  0.010 \\
0.4  &  0.028  &  --2.794   &  0.018 & 0.081 &  --1.817  &  0.049&  0.237  &  --1.057   &  0.105 \\
0.6  &  0.129  &  --0.667  &  0.108  & 0.451 &  0.307  &  0.286 &  0.914  &  0.945   &  0.541 \\
0.7  &  0.030  &   0.446  &  0.241  & 0.920  &  1.340   &  0.590 &  1.521   &  1.876    &  1.009    \\
0.8  &  0.687  &   1.633   &  0.511  & 1.668   &  2.414   &  1.114  &  2.449   &  2.843    &  1.712    \\

\hline
\hline
\end{tabular*}
\label{tblecontact}
\end{table}

More information concerning the radial distribution functions can be found in  table~\ref{tblecontact}, where the results for contact values, $\,g(r=\sigma)$, are collected. Looking for the MC data one can see, that for each of three HCRY fluids the contact value $\,g(r=\sigma)\,$ increases when density increases. However, in the case of the HCRY fluid with $\,z\sigma=5\,$ the radial distribution function at the contact always remains smaller than one, $\,g(r=\sigma)<1\,$, while in the case of two other HCRY fluids with $\,z\sigma=3\,$ and 1.8 this is not the case. From the results presented in table~\ref{tblecontact} it follows that SEXP/FMSA theory results in $\,g(r=\sigma)\,$ values that are always slightly lower than similar MC data, while in the case of the FMSA theory this depends on the density: at low densities FMSA theory results in contact values $\,g(r=\sigma)\,$ that are lower than MC data, but at high densities it reverses, i.e., the MC data for $\,g(r=\sigma)\,$ are smaller than those from the FMSA theory. Thus, there is a crossover density which depends on the range of repulsion (i.e., depends on the value of parameter $\,z\,$). The existence of this crossover density explains why for some HCRY fluid there is a range of densities, where the FMSA theory results in the contact values $\,g(r=\sigma)\,$ that are very close to the MC data.
However, it is evident that on average in all considered cases the SEXP/FMSA approximation improves an agreement between theory and computer simulations, especially, at short distances. At intermediate and large distances, both FMSA and SEXP/FMSA theories are quite similar, and both slightly underestimate the ordering that takes place in the HCRY fluids. In particular, the oscillations of the radial distribution functions predicted by the FMSA and SEXP/FMSA theories show to be slightly weaker than those resulting from computer simulation data.

\begin{figure}[!ht]
  \begin{center}
    \includegraphics[height=5.7cm]{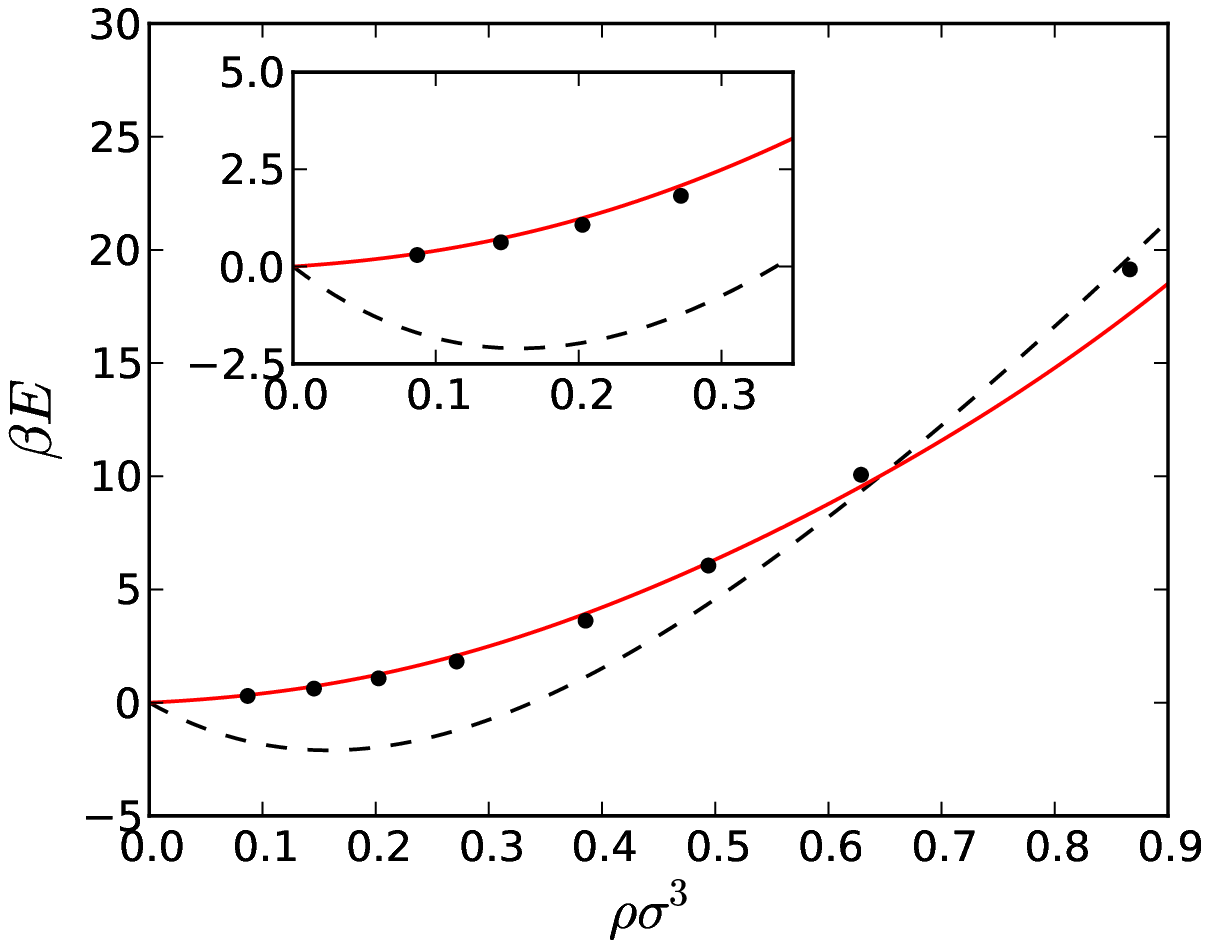}
    \includegraphics[height=5.5cm]{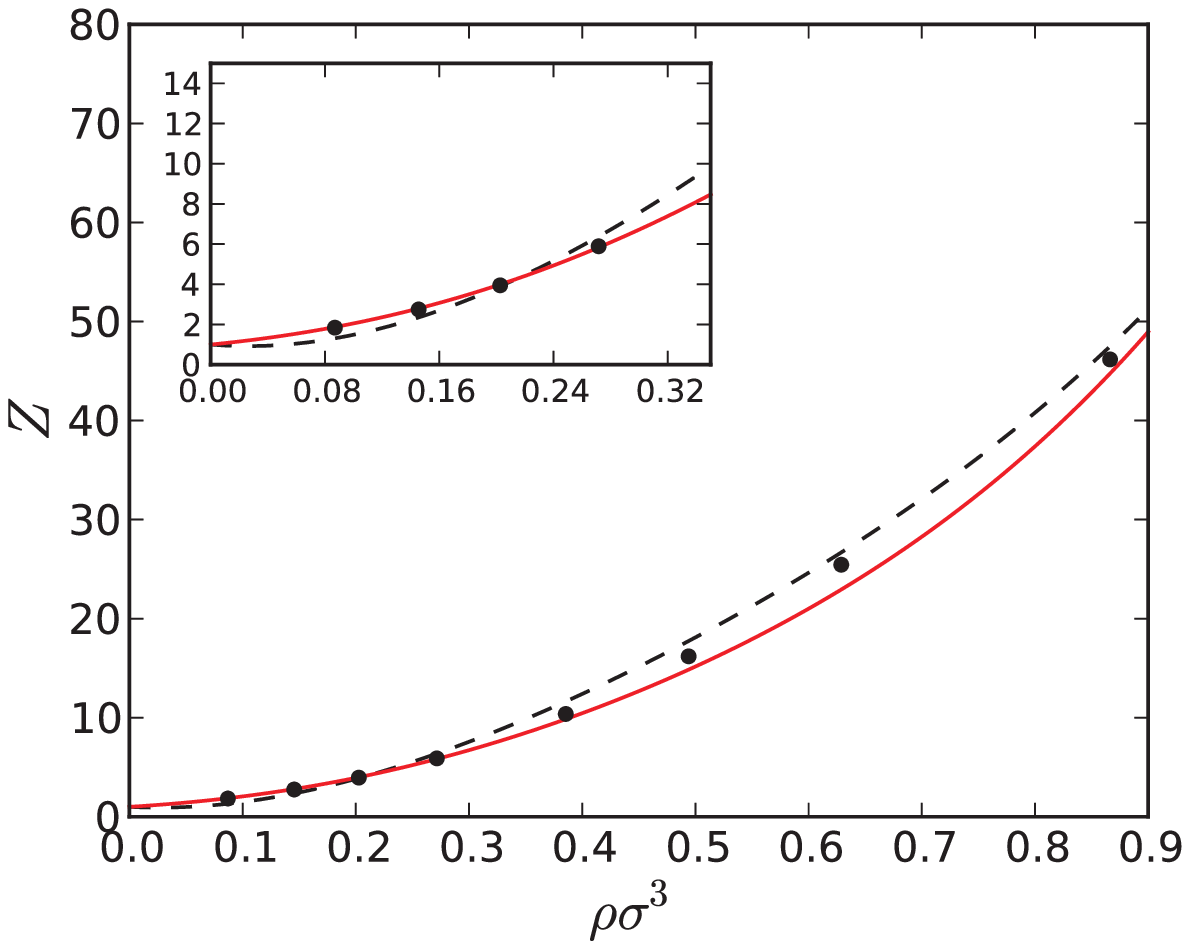}
  \end{center}
  \caption{(Color online) Internal energy (left plot),
and compressibility factor (right plot)
of the hard-core repulsive Yukawa fluid with $z\sigma=5$  at temperature $T^*=0.125$.
Solid black lines denote the results of the FMSA theory, while
the solid red lines mark the results of the SEXP approximation.
Symbols correspond to the MC simulation data.}
  \label{fig:z5ez} 
\end{figure}
\begin{figure}[!h]
  \begin{center}
    \includegraphics[height=5.7cm]{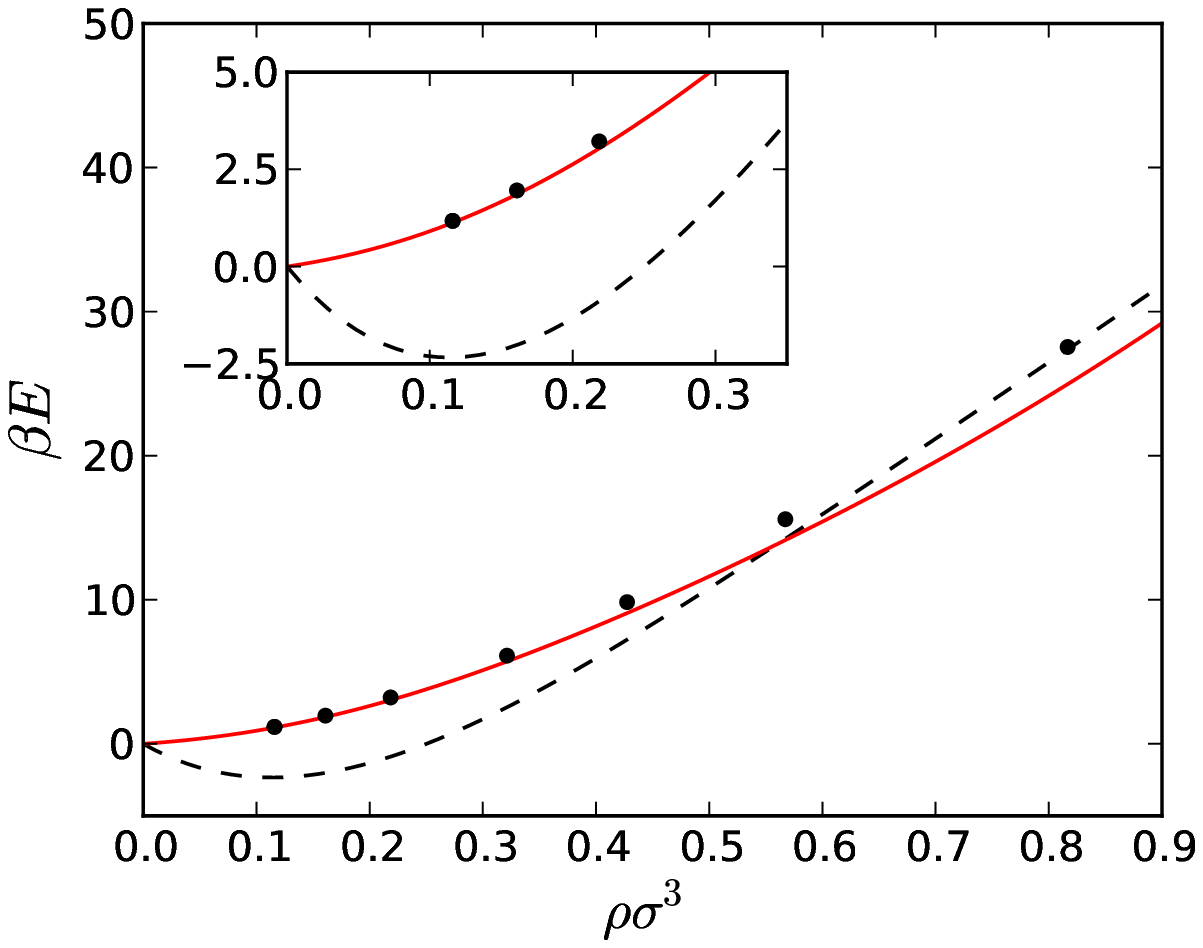}
    \includegraphics[height=5.5cm]{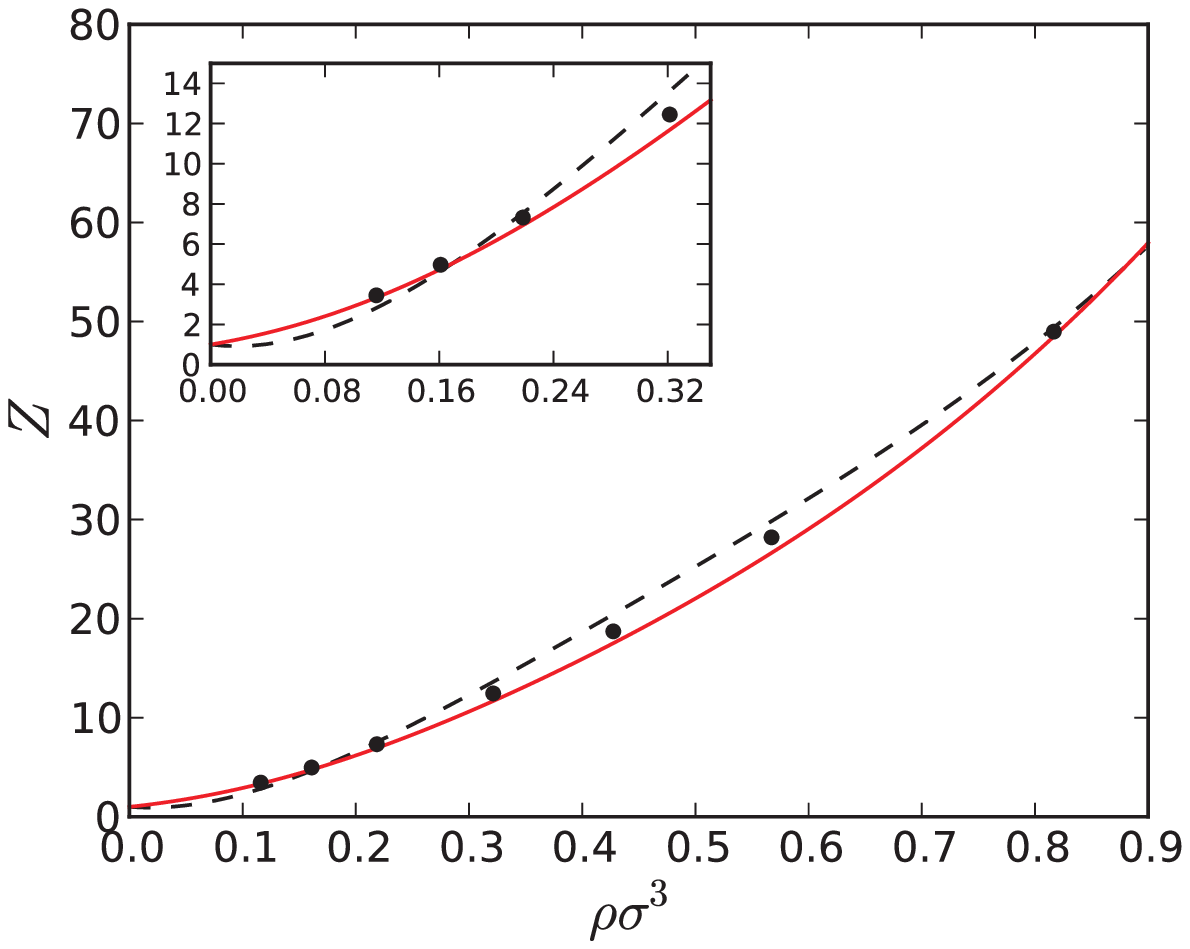}
  \end{center} \vspace{-3mm}
  \caption{(Color online) The same as figure~\ref{fig:z5ez} but for $z\sigma=3$.}
  \label{fig:z3ez}
\end{figure}
\begin{figure}[!h]
  \begin{center}
    \includegraphics[height=5.7cm]{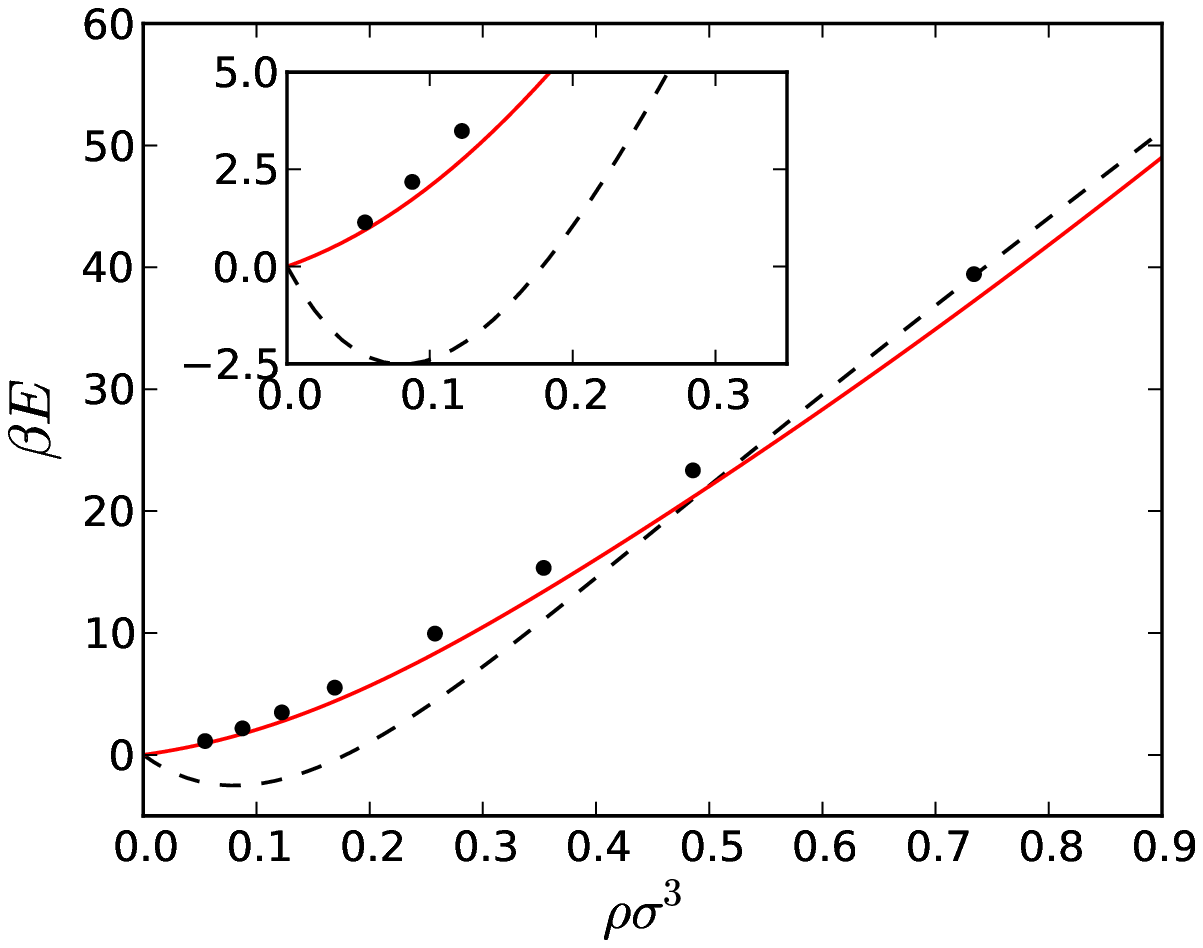}
    \includegraphics[height=5.5cm]{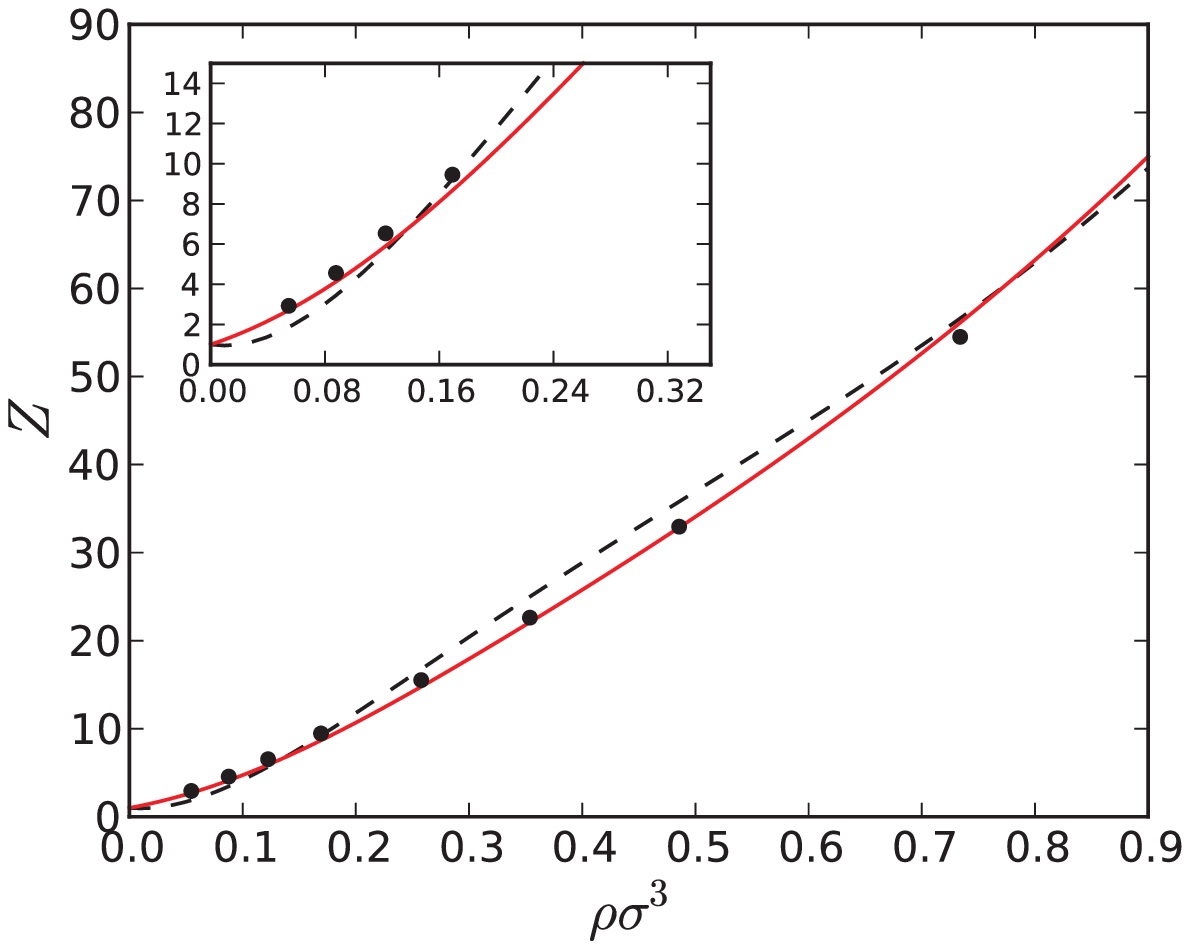}
  \end{center}  \vspace{-3mm}
  \caption{(Color online) The same as figure~\ref{fig:z5ez} but for $z\sigma=1.8$.}
  \label{fig:z18ez}
\end{figure}

\begin{table}[!ht]
\caption{Compressibility factor $\,Z=\beta P/\rho\,$ and internal energy $\,\beta E\,$ of the HCRY fluid with $z\sigma=5$
at reduced temperature $\,T^*=0.125$.}
\vspace{2ex}
\begin{tabular*}{\textwidth}{@{\extracolsep{\fill}}rrrrrrr}
\hline
\hline
 & \multicolumn{3}{c}{$\beta P/\rho$} & \multicolumn{3}{c}{$\beta E$} \\
\cline{2-4}\cline{5-7}
  $\rho\sigma^3$  &  MC  &  FMSA  &  SEXP  &  MC  &  FMSA  &  SEXP \\
\hline
  0.087  &  1.840  &  1.310  &  1.872  &  0.297  &  --1.705  &  0.333 \\
  0.146  &  2.748  &  2.351  &  2.799  &  0.623  &  --2.092  &  0.721 \\
  0.203  &  3.948  &  3.902  &  3.992  &  1.072  &  --1.961  &  1.246 \\
  0.272  &  5.892  &  6.381  &  5.835  &  1.818  &  --1.222  &  2.082 \\
  0.385  &  10.377  &  11.639  &  9.853  &  3.622  &  1.132  &  3.933 \\
  0.494  &  16.195  &  17.744  &  14.855  &  6.056  &  4.371  &  6.182 \\
  0.629  &  25.444  &  26.686  &  22.939  &  10.067  &  9.331  &  9.549 \\
  0.866  &  46.173  &  47.493  &  44.759  &  19.139  &  19.684  &  17.191 \\
\hline
\end{tabular*}
\label{tblez5}
\end{table}

\begin{table}[!ht]
\caption{Compressibility factor $\,Z=\beta P/\rho\,$ and internal energy $\,\beta E\,$ of the HCRY fluid with $\,z\sigma=3\,$
at reduced temperature $\,T^*=0.125$.}
\vspace{2ex}
\begin{tabular*}{\textwidth}{@{\extracolsep{\fill}}rrrrrrr}
\hline
\hline
 & \multicolumn{3}{c}{$\beta P/\rho$} & \multicolumn{3}{c}{$\beta E$} \\
\cline{2-4}\cline{5-7}
  $\rho\sigma^3$  &  MC  &  FMSA  &  SEXP  &  MC  &  FMSA  &  SEXP \\
\hline
 0.116  &  3.450  &  2.820  &  3.338  &  1.175  &  --2.337  &  1.128 \\
 0.161  &  4.971  &  4.660  &  4.757  &  1.957  &  --2.014  &  1.861 \\
 0.219  &  7.321  &  7.556  &  6.938  &  3.218  &  --0.889  &  3.037 \\
 0.321  &  12.448 &  13.623 &  11.688 &  6.119  &  2.538  &  5.717 \\
 0.427  &  18.716 &  20.439 &  17.524 &  9.836  &  7.232  &  9.074 \\
 0.567  &  28.207 &  29.845 &  26.642 &  15.591  &  14.233  &  14.140 \\
 0.817  &  48.982 &  49.399 &  48.497 &  27.545  &  27.373  &  24.965 \\
\hline
\end{tabular*}
\label{tblez3}
\end{table}

\begin{table}[!ht]
\caption{Compressibility factor $\,Z=\beta P/\rho\,$ and internal energy $\,\beta E\,$ of the HCRY fluid with $\,z\sigma=1.8\,$
at reduced temperature $\,T^*=0.125$.}
\vspace{2ex}
\begin{tabular*}{\textwidth}{@{\extracolsep{\fill}}rrrrrrr}
\hline
\hline
& \multicolumn{3}{c}{$\beta P/\rho$} & \multicolumn{3}{c}{$\beta E$} \\
\cline{2-4}\cline{5-7}
  $\rho\sigma^3$  &  MC  &  FMSA  &  SEXP  &  MC  &  FMSA  &  SEXP \\
\hline
  0.055  &  2.929  &  1.887  &  2.727  &  1.136  &  --2.277  &  0.937 \\
 0.088  &  4.561  &  3.455  &  4.133  &  2.176  &  --2.495  &  1.727 \\
  0.122  &  6.534  &  5.685  &  5.910  &  3.486  &  --1.991  &  2.746 \\
 0.169  &  9.456  &  9.244  &  8.679  &  5.517  &  --0.419  &  4.406 \\
 0.258  &  15.521  &  16.759  &  14.773  &  9.951  &  4.415  &  8.331 \\
  0.354  &  22.614  &  24.999  &  22.096  &  15.341  &  11.095  &  13.409 \\
  0.486  &  32.948  &  35.792  &  32.855  &  23.341  &  20.998  &  21.155 \\
 0.734  &  54.504  &  56.609  &  56.117  &  39.437  &  39.313  &  37.236 \\
\hline
\hline
\end{tabular*}
\label{tblez18}
\end{table}

Corresponding results for thermodynamic properties are presented in figures~\ref{fig:z5ez}, \ref{fig:z3ez} and \ref{fig:z18ez} and collected in tables~\ref{tblez5}, \ref{tblez3} and \ref{tblez18}.  In these figures and tables we show a comparison of the density dependence of the internal energy and compressibility factor for all three HCRY fluids, obtained from the FMSA and SEXP/FMSA theories, against the corresponding MC simulation data.  One can see that the above discussed failure of the FMSA theory for the HCRY fluids that results in the negative values of the radial distribution function, is projected into the non-physical negative values of the internal energy for this type of fluids.
Again we can see that SEXP approximation significantly improves theoretical predictions in all considered cases.
This manifests itself in the internal energy for all three HCRY fluids being positive now, and, in general, more accurate than the one that follows from the FMSA theory.
Quite similar, but less pronounced trend is observed for the compressibility factor. The SEXP/FMSA theory performs better everywhere, except at very high densities, where the internal energy
obtained from the FMSA theory seems to be more accurate. However, this still remains questionable, at least for the case of HCRY fluid with $\,z\sigma=5\,$, since according to the phase diagram for this fluid~\cite{AzharJCP2000}, the high densities (e.g., densities $\,\rho\sigma^3>0.7$) are in the vicinity of the transition region or even could be already associated with the solid phase. Similarly, the highest densities for all other HCRY fluids have to be treated with care.

\section{Conclusions}

Summarizing, a simplified exponential SEXP/FMSA theory was employed to describe the thermodynamics of the hard core repulsive Yukawa (HCRY) fluid.
The theory is built up on an analytical solution of the FMSA theory for the radial distribution function due to Tang and Lu~\cite{TangJCP1993}. The main reason that
has pursued us into this project were nonphysical negative values for the radial distribution function of the HCRY fluid  that are predicted by the original FMSA theory at low temperatures. Obviously, this artifact of the FMSA theory will cause problems for the FMSA thermodynamics of the HCRY fluid as well.

The results presented in this study are obtained for the low reduced temperature characterized by $\,T^*=0.125$. Such a low reduced temperature for the HCRY fluids has not been explored within the integral equation theory studies so far. However, there are computer simulation studies of the phase diagram of the HCRY fluid at such and even lower temperatures~\cite{HeinenJCP2011,AzharJCP2000}.
We have shown that the conventional FMSA theory at low density leads to the non-physical negative values of the radial distribution function at short distances and, consequently, to incorrect thermodynamics yielding the negative internal energy, but it seems to be reasonably accurate at higher densities.
By contrast, the exponential approximation based on the FMSA solution significantly improves the theoretical treatment of the HCRY fluids.
The SEXP/FMSA theory proves to be more successful for the HCRY fluids characterized by a more short-ranged repulsive interaction.
In the case of long-ranged repulsive potentials, the exponential approximation performing reasonably well at low densities, tends to predict weaker oscillations of the radial distribution function at higher densities.

\appendix

\section{Coefficients of the inversion of the radial distribution function and their density derivatives}

The other still undefined functions
are as follows:
\begin{eqnarray}
 C\left( n_1,n_2,n_3,x \right) =  \left[\sum_{\alpha=0}^{2}\sum_{i=1}^{n_3}
 \frac{ {x}^{i-1} \re^{t_{\alpha}x}}{\left(i-1\right)!} B\left(n_1,n_2,n_3,i,\alpha\right)
 + \frac{\left( 1+\eta/2 \right)^{n_2}}{\left( 1- \eta \right)^{2n_3}}\delta_{ n_1+n_2 ,3n_3}\right]H\left( x \right)\,, \nonumber
\end{eqnarray}
\begin{eqnarray}
 D\left( n_1,n_2,n_3,y,x \right) &=& \left\{\sum_{\alpha=0}^{2}\sum_{i=1}^{n_3}
 \left[ \re^{-yx}
 - \re^{t_\alpha x} \sum_{j=0}^{i-1} \frac{\left( -1\right)^j\left(t_\alpha+y\right)^j{x}^j}{j!} \right]\right.\nonumber\\
&\times&\left.\frac{\left( -1 \right)^i B\left(n_1,n_2,n_3,i,\alpha\right)}{\left(t_{\alpha}+y\right)^i}
+ \frac{\left( 1+\eta/2 \right)^{n_2}}{\left( 1- \eta \right)^{2n_3}}\re^{-yx}\delta_{ n_1+n_2 ,3n_3}\right\}H\left(x\right)\,, \nonumber
\end{eqnarray}
where
\begin{eqnarray}
B\left( n_1,n_2,n_3,i,\alpha \right) &=&  \frac{1}{\left( 1-\eta \right)^{2n_3}}
 \sum_{k_1=0}^{n_3-i}\sum_{k_2=0}^{n_3-i-k_1}
 \frac{A\left(n_1,n_2,k_1,t_\alpha\right)}{\left(t_\alpha-t_\beta\right)^{n_3+k_2}
 \left(t_\alpha-t_\gamma\right)^{2n_3-i-k_1-k_2}}\nonumber\\
&\times&\frac{\left(n_3-i+k_2\right)!\left(2n_3-1-i-k_1-k_2\right)!}
 {k_1!k_2!\left(n_3-i-k_1-k_2\right)!\left[\left(n_3-1\right)!\right]^2}\left(-1\right)^{n_3-i-k_1}\,,\nonumber
\end{eqnarray}
and
\begin{eqnarray}
 A\left(n_1,n_2,k_1,t_\alpha\right) = \sum_{i=\max\left(k_1-n_1,0\right)}^{n_2}
 \frac{n_2!\left(i+n_1\right)!}{i!\left(n_2-i\right)!\left(i+n_1-k_1\right)!}
 \left(1+\frac12\eta\right)^i\left(1+2\eta\right)^{n_2-i}t_{\alpha}^{n_1+i-k_1}\,.\nonumber
 \label{BandA}
\end{eqnarray}
In these expressions  $H(x)$ is the Heaviside step function, while
$\,t_\alpha\,$ stands for the three ($\,\alpha=0, 1, 2\,$) roots of the equation $\,S\left( t \right)=0\,$ with
 $\,S\left( t \right)\,$ being the polynomial
\begin{equation}
S\left( t\right) =\left( 1-\eta \right) ^{2}t^{3}+6\eta \left(1-\eta\right) t^{2}
                 + 18\eta ^{2}t-12\eta \left( 1+2\eta \right)\,.  \label{S_def}\nonumber
\end{equation}
The roots are given by
\begin{equation}
 t_{\alpha}=\frac{\left[ -2\eta+\left( 2\eta f \right)^{1/3}\left( y_+c^{\alpha}
   + y_-c^{-\alpha} \right) \right]}{1-\eta},\quad \alpha=0,1,2,
   \label{t_alpha}\nonumber
\end{equation}
where
\begin{eqnarray}
f = 3+3\eta-\eta^2\,,\qquad y_{\pm} = \left[ 1\pm \left( 1+\frac{2\eta^4}{f^2} \right)^{1/2}\right]^{1/3}\qquad \text{and}\qquad
c = \re^{2\pi \ri/3}\,.
   \label{f_y_c}\nonumber
\end{eqnarray}
%

Now the derivatives of the $C$ and $D$ functions read
\begin{eqnarray}
  \frac{\partial C\left( n_1,n_2,n_3,x \right)}{\partial\rho}  &=&  \left\{\sum_{\alpha=0}^{2}\sum_{i=1}^{n_3}
  \frac{ {x}^{i-1}}{\left(i-1\right)!}
  \left[
    x\frac{\partial t_{\alpha}}{\partial\rho} \re^{t_{\alpha}x} B\left(n_1,n_2,n_3,i,\alpha\right)
    + \re^{t_{\alpha}x} \frac{\partial B\left(n_1,n_2,n_3,i,\alpha\right)}{\partial \rho}
  \right]\right.
\nonumber\\
 &+&
 \left.\frac{\pi}{6}\left[
 \frac{n_2}{2}\frac{\left( 1+\frac12\eta \right)^{n_2-1}}{\left( 1- \eta \right)^{2n_3}}
+ 2n_3\frac{\left( 1+\frac12\eta \right)^{n_2}}{\left( 1- \eta \right)^{2n_3+1}}
\right]\delta_{ n_1+n_2 ,3n_3}\right\}H\left( x \right),
\nonumber
\end{eqnarray}

\begin{eqnarray}
\frac{\partial D\left( n_1,n_2,n_3,y,x \right)}{\partial\rho} \!\!\!\!\! &=& \!\!\!\!\!
\left\{
\sum_{\alpha=0}^{2}\sum_{i=1}^{n_3}
 \left[
   x \frac{\partial t_{\alpha}}{\partial\rho}
   \re^{t_\alpha x} \sum_{j=0}^{i-1} \frac{\left( -1\right)^j\left(t_\alpha+y\right)^j {x}^j}{j!}
 -   \frac{\partial t_{\alpha}}{\partial\rho}
   \re^{t_\alpha x} \sum_{j=0}^{i-1} \frac{\left( -1\right)^j j
   \left(t_\alpha+y\right)^{j-1}  {x}^j}{j!}
 \right]\right.\nonumber\\
&\times&\!\!\!\!\! \left.\frac{\left( -1 \right)^i B\left(n_1,n_2,n_3,i,\alpha\right)}{\left(t_{\alpha}+y\right)^i}
+\sum_{\alpha=0}^{2}\sum_{i=1}^{n_3}
 \left[\frac{}{} \re^{-yx}
 - \re^{t_\alpha x} \sum_{j=0}^{i-1} \frac{\left( -1\right)^j\left(t_\alpha+y\right)^j {x}^j}{j!}
 \right]\right.\nonumber\\
&\times&\!\!\!\!\! \left( -1 \right)^i\left.
\left[
\frac{\partial B\left(n_1,n_2,n_3,i,\alpha\right)/\partial\rho}{\left(t_{\alpha}+y\right)^i}
- i \frac{\partial t_{\alpha}}{\partial\rho} \frac{B\left(n_1,n_2,n_3,i,\alpha\right)}{\left(t_{\alpha}-y\right)^{i+1}}
\right] \right. \nonumber\\
&+&\!\!\!\!\! \left.
 \frac{n_2\pi\sigma^3}{12}\frac{\left( 1+\eta/2 \right)^{n_2-1}}{\left( 1- \eta \right)^{2n_3}}\re^{-yx}\delta_{ n_1+n_2 ,3n_3}
+ \frac{n_3\pi\sigma^3}{3}\frac{\left( 1+\eta/2 \right)^{n_2}}{\left( 1- \eta \right)^{2n_3+1}}\re^{-yx}\delta_{ n_1+n_2 ,3n_3}
\right\}H\left(x\right)\,, \nonumber
\end{eqnarray}
and
\begin{eqnarray}
\frac{\partial B\left( n_1,n_2,n_3,i,\alpha \right)}{\partial\rho} &=&
\frac{n_3\pi\sigma^3}{3\left(1-\eta\right)}B\left( n_1,n_2,n_3,i,\alpha \right) \nonumber\\
&+& \frac{1}{\left( 1-\eta \right)^{2n_3}}
 \sum_{k_1=0}^{n_3-i}\sum_{k_2=0}^{n_3-i-k_1}
 \left[
 \frac{\partial A\left(n_1,n_2,k_1,t_\alpha\right)/\partial\rho}
 {\left(t_\alpha-t_\beta\right)^{n_3+k_2}\left(t_\alpha-t_\gamma\right)^{2n_3-i-k_1-k_2}} \right.\nonumber\\
 &-&\left.
 \frac{A\left(n_1,n_2,k_1,t_\alpha\right)\left(\frac{\partial t_\alpha}{\partial\rho}-\frac{\partial t_\beta}{\partial \rho}\right)}{\left(t_\alpha-t_\beta\right)^{n_3+k_2+1}\left(t_\alpha-t_\gamma\right)^{2n_3-i-k_1-k_2}}
 -\frac{A\left(n_1,n_2,k_1,t_\alpha\right)\left(\frac{\partial t_\alpha}{\partial\rho}-\frac{\partial t_\gamma}{\partial \rho}\right)}
 {\left(t_\alpha-t_\beta\right)^{n_3+k_2}\left(t_\alpha-t_\gamma\right)^{2n_3-i-k_1-k_2+1}}
 \right]
\nonumber\\
&\times&\frac{\left(n_3-i+k_2\right)!\left(2n_3-1-i-k_1-k_2\right)!}
 {k_1!k_2!\left(n_3-i-k_1-k_2\right)!\left[\left(n_3-1\right)!\right]^2}\left(-1\right)^{n_3-i-k_1}\,,\nonumber
\end{eqnarray}
with
\begin{eqnarray}
\frac{\partial A\left(n_1,n_2,k_1,t_\alpha\right)}{\partial\rho} &=&  \sum_{i=\max\left(k_1-n_1,0\right)}^{n_2}
 \frac{n_2!\left(i+n_1\right)!}{i!\left(n_2-i\right)!\left(i+n_1-k_1\right)!}
 \left[
 \frac{i\pi\sigma^3}{12}\left(1+\frac12\eta\right)^{i-1}\left(1+2\eta\right)^{n_2-i}t_{\alpha}^{n_1+i-k_1}
 \right.\nonumber\\
&+&\left.\frac{\left(n_2-i\right)\pi\sigma^3}{3}\left(1+\frac12\eta\right)^{i}\left(1+2\eta\right)^{n_2-i-1}t_{\alpha}^{n_1+i-k_1}\right.\nonumber\\
&+&\left.\left(n_1+i-k_1\right)\frac{\partial t_{\alpha}}{\partial\rho}\left(1+\frac12\eta\right)^{i}\left(1+2\eta\right)^{n_2-i}t_{\alpha}^{n_1+i-k_1-1}\nonumber
 \right].
 \nonumber
\end{eqnarray}
In these equation
%
\begin{equation}
 \frac{\partial t_{\alpha}}{\partial \rho}=
 \frac{\pi\sigma^3}{6} \left\{ \frac{t_\alpha}{1-\eta}
 +  \frac{1}{1-\eta}
 \left[-2 + \left(2f+2\eta \frac{\partial f}{\partial\eta}\right)
 \frac{\left(y_+c^{j} + y_-c^{-j}\right)}{3 \left(2\eta f\right)^{\frac{2}{3}}}
            + \left(2\eta f\right)^{\frac{1}{3}}
            \left(\frac{\partial y_+}{\partial\eta}c^j + \frac{\partial y_-}{\partial\eta}c^{-j}\right)
    \right]
 \right\}, \nonumber\label{dt_deta}
\end{equation}
where
\begin{eqnarray}
 \frac{\partial y_{\pm}}{\partial \eta} =
 \pm \frac{1}{6 y_{\pm}^2 \left(1 + {2\eta^4}/{f^2}\right)^{\frac{1}{2}}}
 {\left( \frac{8\eta^3}{f^2} - \frac{4\eta^4}{f^3} \frac{\partial f}{\partial \eta} \right)} \qquad \text{and}\qquad
 \frac{\partial f}{\partial \eta} = 3-2\eta\label{df_deta}\,.
 \label{dypm_deta}\nonumber
\end{eqnarray}
%



\ukrainianpart

\title{╤яЁю∙хэх хъёяюэхэЄэх эрсышцхээ  фы  ЄхЁьюфшэрь│ъш ЄтхЁфюёЇхЁэюую яышэє ч ■ърт│тё№ъшь т│ф°Єютїєтрээ ь}
\author{╤. ├ыє°ръ\refaddr{VU,IP}, └. ╥Ёюїшьўєъ\refaddr{IP}}
\addresses{
\addr{VU} ╘ръєы№ЄхЄ ї│ь│ўэю┐ Єр с│юьюыхъєы Ёэю┐  │эцхэхЁ│┐, ╙э│тхЁёшЄхЄ ┬рэфхЁс│ы№фр, ═х°т│ы, ╥хээхё│, ╤╪└
\addr{IP} ▓эёЄшЄєЄ Ї│чшъш ъюэфхэёютрэшї ёшёЄхь, ═└═ ╙ъЁр┐эш, 79011 ╦№т│т, ╙ъЁр┐эр
}

\newpage

\makeukrtitle

\begin{abstract}
\tolerance=3000%
┼ъёяюэхэЄэх эрсышцхээ  эр юёэют│ ёхЁхфэ№ю-ёЇхЁшўэюую эрсышцхээ  яхЁ°юую яюЁ фъє (FMSA) чрёЄюёютє║Є№ё  фю тштўхээ  ёЄЁєъЄєЁш Єр ЄхЁьюфшэрь│ъш ЄтхЁфюёЇхЁэюую яышэє ч ■ърт│тё№ъшь т│ф°Єютїєтрээ ь. ╟ряЁюяюэютрэр ЄхюЁ│  ъюЁшёЄє║Є№ё  ЇръЄюь, ∙ю  │ёэє║ рэры│Єшўэшщ Ёючт' чюъ FMSA, р чтшўрщэх хъёяюэхэЄэх эрсышцхээ  ёєЄЄ║тю щюую яюъЁр∙є║. ╪ы їюь яюЁ│тэ ээ  ч ъюья'■ЄхЁэшь хъёяхЁшьхэЄюь яюърчрэю, ∙ю т юсырёЄ│ ьрышї уєёЄшэ Єр эшч№ъшї ЄхьяхЁрЄєЁ, Єрь фх ЄхюЁ│  ёхЁхфэ№юую яюы  эх яЁрЎ■║, хъёяюэхэЄэх эрсышцхээ  яЁштюфшЄ№ фю ёєЄЄ║тюую тфюёъюэрыхээ  ЄхюЁхЄшўэшї яхЁхфсрўхэ№.
\keywords яышэ ч ■ърт│тё№ъшь т│ф°Єютїєтрээ ь, ёхЁхфэ№ю-ёЇхЁшўэх эрсышцхээ  яхЁ°юую яюЁ фъє,
хъёяюэхэЄэх эрсышцхээ , ╠юэЄх ╩рЁыю ьюфхы■трээ 
\end{abstract}

\end{document}